%% file: main.tex
\documentclass[12pt]{article}

\usepackage{authblk}

\usepackage{amsmath,amssymb,amsfonts,amsthm}
\usepackage{siunitx}
\usepackage{algorithmic}
\usepackage{graphicx}
\usepackage{textcomp}
\usepackage{xcolor}
\usepackage{bbm}
\usepackage{optidef}
\usepackage{enumerate}
\usepackage{hyperref}
\usepackage[capitalise]{cleveref}
\crefrangeformat{equation}{(#3#1#4)--(#5#2#6)}
\usepackage{subcaption}

\usepackage{chngcntr}

\usepackage{xspace}

\usepackage{tikz}
\usepackage{circuitikz}
\usepackage{pgfplots, pgfplotstable}
\usepgfplotslibrary{groupplots}
\pgfplotsset{compat=newest}
\usetikzlibrary{shapes,arrows,positioning,calc}
\usetikzlibrary{arrows.meta}
\begin{document}

% writing commands
\newcommand{\todo}[1]{\textcolor{orange}{(#1)}}

% mathematical commands
\newcommand{\onenorm}[1]{\lVert{#1}\rVert_{1}}
\newcommand{\twonorm}[1]{\lVert{#1}\rVert_{2}}
\newcommand{\inv}{^{-1}}
\newcommand{\ones}{\mathbf{1}}
\newcommand{\zeros}{\mathbf{0}}
\newcommand{\R}{\mathbb{R}}
\newcommand{\ramp}[3]{\mathrm{ramp}_{#2}^{#3}\left({#1}\right)}
% graph notation
\newcommand{\G}{\mathcal{G}}
\newcommand{\Nn}{\mathcal{N}}
\newcommand{\E}{\mathcal{E}}
\newcommand{\Vv}{\mathcal{V}}
\newcommand{\pathL}{\mathcal{L}}

%units
\newcommand{\qunit}{l$/$min\xspace}
\newcommand{\punit}{mH$2$O\xspace}

% environments
\newtheorem{assumption}{Assumption}

% ===============================================================
% FRONT PAGE
% ===============================================================
\title{Hydraulic Parameter Estimation for District Heating Based on Laboratory Experiments}

\author[1]{Felix Agner\thanks{Corresponding author: \href{mailto:felix.agner@control.lth.se}{felix.agner@control.lth.se}.}}
\author[2,3]{Christian Møller Jensen}
\author[1]{Anders Rantzer}
\author[2,3]{Carsten Skovmose Kallesøe}
\author[2]{Rafal Wisniewski}

\affil[1]{Department of Automatic Control, Lund University, Lund, Sweden}

\affil[2]{Department of Electronic Systems, Aalborg University, Aalborg,Denmark}

\affil[3]{Grundfos Holdings A/S, Bjerringbro, Denmark}

\date{} % This will remove the date
\maketitle

\begin{abstract}
In this paper we consider calibration of hydraulic models for district heating systems based on operational data. We extend previous theoretical work on the topic to handle real-world complications, namely unknown valve characteristics and hysteresis. We generate two datasets in the Smart Water Infrastructure laboratory in Aalborg, Denmark, on which we evaluate the proposed procedure. In the first data set the system is controlled in such a way to excite all operational modes in terms of combinations of valve set-points. Here the best performing model predicted volume flow rates within roughly 5 and 10 \% deviation from the mean volume flow rate for the consumer with the highest and lowest mean volume flow rates respectively. This performance was met in the majority of the operational region. In the second data set, the system was controlled in order to mimic real load curves. The model trained on this data set performed similarly well when evaluated on data in the operational range represented in the training data. However, the model performance deteriorated when evaluated on data which was not represented in the training data.
\end{abstract}

\input{paper}

\end{document}

%% file: paper.tex
\section{Introduction}

The energy systems of the near-future face many challenges such as higher penetration of renewable energy production, and large and fluctuating energy demand. To face these challenges, a promising outlook is to introduce a stronger coupling between smart girds, smart thermal grids and smart gas grids. In this proposed architecture, the emerging 4th generation of district heating systems, characterized by decentralized heat production, significant integration with smart power and gas grids, and low distribution temperatures, is likely to take a central role \cite{lund_4th_2014}. Operating these new systems come with many challenges related to control and operation \cite{vandermeulen_controlling_2018}. To face these new challenges, smart systems and techniques for modelling and simulating the systems will be key. While modelling and simulating the thermodynamics of district heating system for operational optimization now has a now rather long tradition \cite{benonysson_operational_1995}, recent works have also started to delve deeper into the hydraulics of district heating systems. Hydraulic models are useful for simulating the hydraulic state of district heating networks \cite{vesterlund_simulation_2016, denarie_dynamical_2023}, and such simulations can be used to operate the systems in an optimal fashion \cite{guelpa_optimal_2016, WANG201883, agner_combating_2022, zheng_thermo-hydraulic_2023}. In essence, hydraulic models describe the relationship between pressures and volume flow rates for hydraulic system components such as valves and pipes. Traditionally, such models have consisted of fully white-box models. For pipes this implies models based on friction coefficients and pipe dimensions. For valves, it has implied manufacturer-provided characteristic curves. While the emerging energy systems pose many tough engineering problems, they also open the door to improving these models by tuning them based on operational data. This is due to the expected increase of smart metering in modern systems \cite{vandermeulen_controlling_2018}. This data-driven approach holds many advantages compared to traditional methods. Data driven models can be updated to better fit the observed behavior of a component under operation and they can be adapted over time to compensate for degradation. On the contrary, gathering and maintaining correct information to build traditional white-box models for hydraulic components, e.g. pipe dimensions and friction coefficients, can be tedious and error-prone. While one could consider fully black-box models for specific purposes, there are advantages to maintaining a component-based grey-box model structure. This approach maintains physical interpretability and allows the model to be used in applications other than simulation, such as fault detection \cite{bahlawan_detection_2022}. 

The notion of data-driven model calibration has been studied longer in the field of water distribution networks, as reviewed in \cite{quovadisWDS}, than in the field of district heating. However, district heating systems typically have a distinct structure from water distribution networks, which can be exploited in modeling efforts: District heating systems form closed hydraulic circuits, whereas water distribution networks end in open outlets at the points of consumption. Several works aimed particularly at modeling in the district heating domain have been published in recent years. The authors of \cite{WANG201883} performed pipe resistance estimation and subsequently used the identified model to demonstrate effectiveness of an advanced control strategy in simulation. They estimated the resistance of all pipes in the supply-and-return-lines separately, based on measurement of pressure at the supply-and-return connections of each substation. In \cite{DH_pipe_hydraulic_estimation}, the authors investigated the same strategy while demonstrating a way to account for meshed structures in the system by successively closing sets of loop-generating pipes. In \cite{parameter_estimation_fewer_measurements}, yet a third simulation study with this approach was performed, where the authors now focused on reducing the number of required measurements in the system. In \cite{AGNER2023hydraulic_parameter_estimation}, an extension of the framework was proposed where estimation of valve characteristics was included through exploiting the closed hydraulic circuit structure of district heating networks. A common trend among these works is that they are based on mathematical analysis of the networks along with simulations. Zheng et. al. on the other hand perform pipe parameter identification and compare their model to real operational data from a real system in Tianjin, China \cite{ZHENG2024_digital_twin}. This is the first and only work to the authors' knowledge that performs a study on real data. However, this study is restricted to models of pipe parameters, excluding valve models. They are also unable to share their data with the scientific community. Studying data from real systems is highly important, as many issues which are not visible in simulation can emerge. Measurement noise causes an obvious issue, but also problems based on unmodelled properties of the system. A clear example of such an aspect which is not covered in previous literature on hydraulic district heating model calibration, but which is a well-known issue to engineers working on physical systems, is valve hysteresis \cite{tore_practice}.

In this work, we perform experiments in a laboratory setup as a step towards bridging the gap between the theoretical literature and real application. The study was performed in the Smart Water Infrastructure Laboratory (SWIL) at Aalborg University \cite{smart_water_lab}. We present the following contributions.
\begin{enumerate}
    \item We build a test-bed simulating the hydraulic properties of four buildings in a line-structured district heating system.
    \item We generate two data-sets. Firstly a data-set designed to excite all operational modes of the system. Secondly a data-set based on real district heating load curves. These data sets are open and available for future work.
    \item We extend the modeling approach presented in \cite{AGNER2023hydraulic_parameter_estimation} to handle real world complications. Namely, valve characteristics are not known a priori, and valves are subject to hysteresis.
\end{enumerate}

The paper is organized in the following way. We formally introduce the problem and the considered parameter estimation framework in Section \ref{sec:background}. We then go into detail regarding the specific model structures we consider in Section \ref{sec:model structures}. The experimental setup is described in Section \ref{sec:experiment}. We present and discuss the results of our study in Section \ref{sec:results}. We conclude the paper in Section \ref{sec:conclusion} and present topics for future work in Section \ref{sec:future work}.

\subsection{Notation}
We denote volume flow rates (\qunit) by the variable $q$, pressures (\punit) by the variable $p$ and valve set-points by the variable $v$. We use subscripts to connect the measured values to a specific component, e.g. $q_i$ is the volume flow rate measured in the component of index $i$. For a matrix $M$ we denote $M_{i,j}$ to be the entry of $M$ at row $i$ and column $j$.
We denote the ramp function as
\begin{equation}
    \ramp{x}{a}{b} = \begin{cases}
        0, & \text{if } x \leq a \\
        (x-a)/(b-a), & \text{if } a < x \leq b \\
        1, & \text{else}
    \end{cases}
    \label{eq:ramp function}
\end{equation}
where we assume that $b > a$. We denote a column vector of all 1's of size $n$ as $\ones_{n}$ and an $n$ by $m$ matrix of all zeros as $\zeros_{n \times m}$.

\section{Hydraulic Parameter Estimation} \label{sec:background}
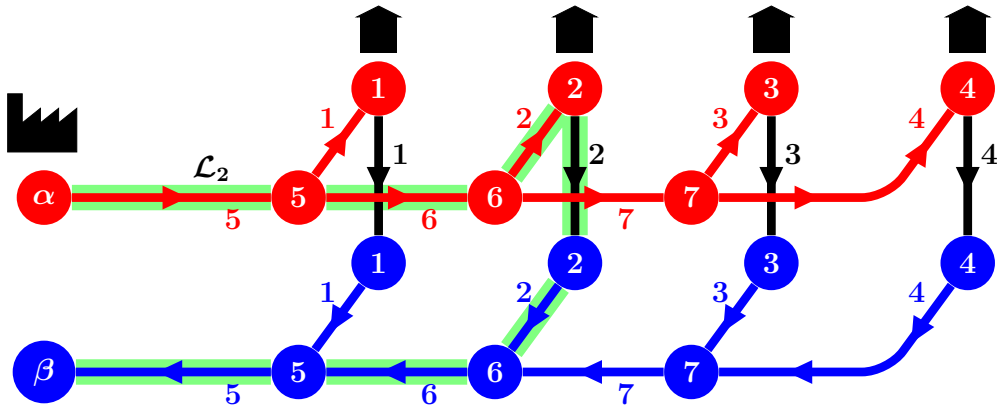
\begin{figure}
    \centering
    \resizebox{\linewidth}{!}{
    \input{fig/grid-schematic}
    }
    \caption{A schematic example of a tree-structured district heating system. The nodes $\Nn = \{\alpha, 1, \dots, 7 \}$ are connected by edges $\E = \{1,\dots,7 \}$. The supply-and-return-networks are symmetrical except for their edges having opposite direction. Valves $\Vv = \{ 1, \dots, 4\}$ connect the supply and return lines. Root nodes $\alpha$ and $\beta$ in the supply-and-return networks respectively connect the branch either to a production facility, or a greater part of the grid. The path $\mathcal{L}_2 = \{2, 5, 6 \}$ highlighted in green contains the edges leading from $\alpha$ to $\beta$ via valve 2.}
    \label{fig:graph-example}
\end{figure}
We consider the problem of finding parameters for hydraulic models of district heating networks, based on operational data corresponding to load conditions measured at times $t = 1, \dots , T$. We restrict ourselves to networks that have a tree-shaped structure, an example of which can be seen in \cref{fig:graph-example}. We model the supply-side network as a structured graph $\G = (\E, \Nn)$ which we assume to have a tree-structure. $\E$ is the set of edges which represent pipes, and $\Nn$ is the set of nodes where these pipes connect. This graph-based perspective is a standard way of representing district heating hydraulics \cite{dePersis-networks}. The supply-and-return networks are assumed to be symmetrical and therefore the return network can be represented by the same graph $\G$, with two exceptions. Firstly, we assume that the edges have opposite direction in the supply and return networks. Hence as the water flows out through the supply network and returns via the return network, the sign of the flow rates through these edges will be the same. Secondly, we consider one node in the graph to be the root of the tree, which we denote $\alpha$ in the supply network and $\beta$ in the return network. This root can represent e.g. a production facility, or a connection to a larger grid to which the considered network is a small sub-grid. A subset $\Vv \subset \Nn$ of the nodes correspond to consumers. They connect the supply and return lines via control valves. In the considered example, $V = \{1,2,3,4 \}$. In a real setting, a single consumer may in fact be represented by two (or more) control valves, as domestic hot water and space heating are typically hydraulically separated \cite[p.365]{thebible}. For each valve $i \in \Vv$, there is a set $\mathcal{L}_i$ of edges corresponding to a path from $\alpha$ to $\beta$ through valve $i$. For instance, $\mathcal{L}_2 = \{2,5,6\}$ in our example as highlighted in \cref{fig:graph-example}.

For the considered parameter estimation protocol, we assume that we can measure the root pressure difference $\Delta p_0 = p_\alpha - p_\beta$. We additionally assume that we can log the control valve set-points $v_i$ for each consumer and the volume flow rates $q_i$ through each valve $i \in \Vv$.

The networks are also assumed to have the following property.
\begin{assumption}
    For all $i \in \Nn$, either $i \in \Vv$ or $i$ is connected to at least 3 edges.
    \label{ass:nodes have many pipes}
\end{assumption}
In practice, this assumption means that no pipes in the model are placed in direct series. Any physical interconnection of pipes in direct series would be represented by one single, equivalent pipe in the model. For instance, it would be reasonable to expect that a consumer substation has a physical interconnection of several pipe sections in series with the control valve and a heat exchanger. The heat exchanger and the pipes would then be concatenated into one single pipe in the model.

With the considered model in mind, we have two types of hydraulic components; \textit{pipes} and \textit{valves}. For each such component, we aim to parameterize a model which maps the volume flow rate through the component to the pressure difference. 

\subsection{Pipe Parameterization}
For pipes, we consider models stemming from the Darcy-Weissbach equation. If the pipe corresponds to edge $i$, which leads from node $j$ to node $k$, the relation between the pressure loss $\Delta p_i = p_j - p_k$ and the volume flow rate $q_i$ along the direction of the pipe is given by
\begin{equation}
    \Delta p_i = s_i q_i |q_i|^\gamma.
    \label{eq:pipe base model}
\end{equation}
Here $s_i$ is the model parameter which our framework aims to estimate from data, typically referred to as the pipes hydraulic resistance. $\gamma$ is an exponent which in the literature typically takes the value 1 or slightly lower e.g. 0.87. We limit the scope of this study to consider the assumption $\gamma = 1$. We make the following assumption regarding the parameters $s_i$.

\begin{assumption}
    The pipe resistance $s_i$ is equal for the corresponding supply-and-return-lines.
    \label{ass:equal supply and return}
\end{assumption}

It was shown in \cite{AGNER2023hydraulic_parameter_estimation} that this assumption along with Assumption \ref{ass:nodes have many pipes} is sufficient for unique identification of model parameters. In practice, it is unlikely in that this assumption holds exactly due to slight differences in the pipes. However, we cannot expect to find unique parameters for these pipes with our methodology, given that we only measure pressure at two points in the network ($\alpha$ and $\beta$). However, for many model applications such as estimating volume flow rates (see Appendix \ref{app: forward estimation}), only the sum of the parameters for the supply-and-return-lines is necessary.

\subsection{Valve Parameterization}
\begin{figure}
    \centering
    \input{fig/valve_curves}
    \caption{Examples of valve characteristics $k_i(v_i)$.}
    \label{fig:valve curve examples}
\end{figure}
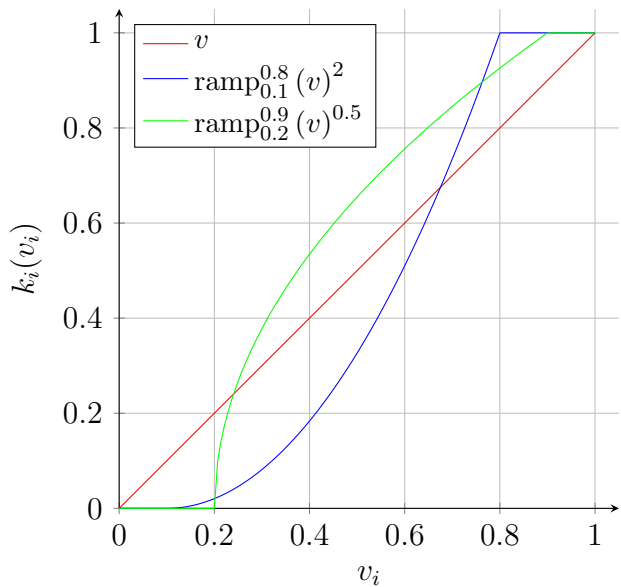
The second and final hydraulic model component type we consider is valves. For valve $i$ which connects node $i$ in the supply network to the return network, $\Delta p_i$ is the pressure difference between these nodes. $\Delta p_i$ is given by the volume flow rate $q_i$ and the valve set-point $v_i$ which varies from fully closed ($v_i = 0$) to fully open ($v_i=1$) in the following way.
\begin{equation}
    q_i = K_{v,i}k_i(v_i)\sqrt{\Delta p_i} \implies \Delta p_i = \frac{1}{K_{v,i}^2k_i(v_i)^2}q_i^2
    \label{eq:valve characteristics}
\end{equation}
where $K_{v,i}$ represents the hydraulic admittance of the valve when it is fully open. This is often referred to as the flow coefficient of the valve, or the $k_{vs}$ or $c_{vs}$ value in Nordic or English literature respectively \cite[p. 399]{thebible}. $k_i$ represents the valve characteristics, i.e. the mapping between the valve set-point and the valve admittance. The valve characteristic function may be linear, e.g. $k_i(v_i) = v_i$. It may however also have other shapes. One common valve characteristic is quick opening, where at low set-points, a small  increase in valve set-point will lead to a large increase in flow-through. Another common characteristic is equal percentage which is in a sense the opposite of quick opening. Here it is expected that the valve needs to provide a larger change in volume flow rate in response to small changes in set-point at higher levels of opening \cite[pp. 82-83]{tore_practice}. Additionally, one might consider valve characteristics where the valve fully closes at a value $v > 0$ and fully opens at a value $v < 1$, to ensure that the full capacity of the valve is used at high set-points, and that the valve is fully closed at near-0 set-points. \cref{fig:valve curve examples} shows three examples of $k_i(\cdot)$, corresponding to a linear valve ($k_i(v_i) = v_i$), an equal percentage valve which fully closes at 0.1 and fully opens at 0.8 ($k_i(v_i) = \ramp{v_i}{0.1}{0.8}^2$), and a quick-opening valve which fully closes at 0.2 and fully opens at 0.9 ($k_i(v_i) = \ramp{v_i}{0.2}{0.9}^{0.5}$). As we cannot necessarily know the valve characteristics of all of the valves we aim to parameterize, we propose a linear combination of typical characteristics, aimed to capture \eqref{eq:valve characteristics}:
\begin{equation}
    \Delta p_i = \left(\frac{\theta_{i,1}}{k_1(v_i)^2} + \frac{\theta_{i,2}}{k_2(v_i)^2} + \dots + \frac{\theta_{i,K}}{k_K(v_i)^2} \right)q_i^2.
    \label{eq:valve model}
\end{equation}
Here $k_k(\cdot)$, $k=1,\dots,K$ correspond to different possible valve characteristics and $\theta_{i,k}$ corresponds to a tunable parameter for valve $i$ in relation to the valve characteristic $k_k(\cdot)$. Somehow $1/k_k(\cdot)^2)$ are basis of all possible characteristics of a valve in a $K$-dimensional space. Hence while the actual valve characteristic may not be perfectly captured by any of the functions $k_k(\cdot)$, a linear combination of them may be able to replicate the actual behavior of the valve.

\subsection{Parameter Estimation}
To estimate the values of $s_i$ (for pipes) and $\theta_{i,k}$ (for valves), we assume that we have access to data from $T$ different steady state load conditions. We first consider one such load condition at time $t$. Here we measure the pressure difference at the tree root $\Delta p_0(t) = p_\alpha(t) - p_\beta(t)$, the valve flow rates $q_i(t)$, and valve set-points $v_i(t)$ where $i \in \Vv$. 

We can use the valve volume flow rates $q_i(t)$ to calculate the volume flow rates in all of the pipes. To do so, we first define the basic incidence matrix $B \in \R^{(n_\Vv-1) \times n_\E}$ for our considered graph representation of the network $\G$. Here $n_\E$ and $n_\Vv$ are the number of edges and nodes respectively. The elements of $B$ are given by
\begin{equation}
    B_{i,j} = \begin{cases}
        1, & \text{if edge } j \text{ leads to node } i, \\
        -1, & \text{if edge } j \text{ leads from node } i, \\
        0, & \text{else}.
    \end{cases}
\end{equation}
We omit the row corresponding to the root node $\alpha$ or $\beta$, which together with the tree-structure of $\G$ ensures the invertibility of $B$. Hence we can find the vector $q_\E$ corresponding to the vector of all flow rates in $\E$ as
\begin{equation}
    q_\E(t) = B\inv q_\Nn(t)
\end{equation}
where $q_\Nn(t) \in \R^{n_\Nn - 1}$ corresponds to the total volume flow rate out of each node $i \in \Nn \setminus \alpha$. I.e.
\begin{equation}
    q_{\Nn,i}(t) = \begin{cases}
        q_i(t), & \text{if } i\in \Vv, \\
        0, & \text{else}.
    \end{cases}
\end{equation}
With this information, the volume flow rates through all pipes and valves are known. We can now consider one $i \in \Vv$, and the corresponding path $\pathL_i$ from $\alpha$ to $\beta$ through valve $i$. It must hold that the total pressure difference $\Delta p_0(t)$ between $\alpha$ and $\beta$ corresponds to the pressure drop along each traversed component in $\pathL_i$, i.e.
\begin{align}
    & \Delta p_0(t) = \Delta p_i(t) \nonumber + 2\sum_{j \in \pathL_i} \Delta p_j(t) \\
    &= \scalebox{.98}{$\left(\frac{\theta_{i,1}}{k_1(v_i(t))^2} + \frac{\theta_{i,2}}{k_2(v_i(t))^2} + \dots + \frac{\theta_{i,K}}{k_K(v_i(t))^2} \right)q_i(t)^2 + 2\sum_{j \in \pathL_i} s_j q_j(t)|q_j(t)|.$}
\end{align}
Here $\Delta p_j(t)$ is equal for the edges in the supply-and-return networks due to Assumption \ref{ass:equal supply and return} and symmetry, yielding the factor 2 in front of the sum. We can use all paths $\pathL_i$ where $i = 1, \dots, n_\Vv$ to formulate a system of equations
\begin{equation}
    \ones_{n_\Vv}\Delta p_0(t) = 
        \scalebox{.95}{$\begin{bmatrix}
        F_1(t) & \zeros_{1 \times K} & \hdots & \zeros_{1 \times K} \\
        \zeros_{1 \times K} & F_2(t) & \hdots & \zeros_{1 \times K} \\
        \vdots & \vdots & \ddots & \vdots \\
        \zeros_{1 \times K} & \zeros_{1 \times K} & \hdots & F_{n_\Vv}(t)
    \end{bmatrix}\begin{bmatrix}
        \theta_1 \\ \theta_2 \\ \vdots \\ \theta_{n_\Vv}
    \end{bmatrix} $} + G(t)s= F(t)\theta + G(t)s.
\end{equation}
Here $G(t) \in \R^{n_\pathL \times n_\E}$ is a matrix defined as 
\begin{equation}
    G_{i,j} = \begin{cases}
        2 q_j(t)|q_j(t)| & \text{if edge } j\in \pathL_i, \\
        0 & \text{else.}
    \end{cases}
\end{equation}
The vectors $\theta_i$ for $i = 1,\dots, n_\Vv$ gather the parameters $\theta_{i,k}$ for $k= 1,\dots,K$, and $F_i(t) \in \R^{1 \times K}$ is given by
\begin{equation}
    F_{i}(t) = \begin{bmatrix}
        \frac{q_i(t)^2}{k_1(v_i(t))^2} & \frac{q_i(t)^2}{k_2(v_i(t))^2} & \hdots & \frac{q_i(t)^2}{k_K(v_i(t))^2}
    \end{bmatrix}.
\end{equation}
We now move from considering only one load condition at time $t$ to concatenating the data from all $T$ load conditions into one large system of equations
\begin{equation}
    \Phi \begin{bmatrix} \theta \\ s \end{bmatrix} = y
    \label{eq:matrix equation}
\end{equation}
where
\begin{equation}
    \Phi = \begin{bmatrix}
        F(1) & G(1) \\
        F(2) & G(2) \\
        \vdots & \vdots \\
        F(T) & G(T)
    \end{bmatrix}, \quad y = \begin{bmatrix}
        \ones_{n_\Vv}\Delta p_0(1)\\
        \ones_{n_\Vv}\Delta p_0(2)\\
        \vdots \\
        \ones_{n_\Vv}\Delta p_0(T)
    \end{bmatrix}.
\end{equation}
This way $\Phi$ is a data-matrix constructed by measuring flow rates and valve set-points and $y$ consists of measured root pressure measurements. Once $\Phi$ and $y$ are constructed, there are many ways to choose $s$ and $\theta$ to fit the observed data. We choose to fit the parameters numerically as the solution to
\begin{mini!} 
    {s,\theta}{\onenorm{\Phi \begin{bmatrix} s \\ \theta \end{bmatrix} - y} }
    {\label{eq:optimal problem}}{\label{eq:optimal cost}}
    \addConstraint{s \geq 0} \label{eq:constraint s positive}
    \addConstraint{\theta \geq 0} \label{eq:constraint theta positive}.
\end{mini!}
The choice of using $\onenorm{\cdot}$ in \eqref{eq:optimal cost} is based on the general robustness to outliers typically provided by this cost function. The positivity constraints \crefrange{eq:constraint s positive}{eq:constraint theta positive} enforce physical feasibility of the model.  We solved \eqref{eq:optimal problem} using the python toolbox \texttt{CVXPY} \cite{diamond2016cvxpy, agrawal2018rewriting} using the \texttt{SciPy} solver. A natural extension of \eqref{eq:optimal problem} is to introduce parameter regularization to \eqref{eq:optimal cost}. This would reduce the risk of overfitting and promote sparseness in the model depending on the choice of regularization employed. Our investigation has however showed that the models we found are already sparse, and therefore leave an investigation of the impact of regularization outside the scope of this paper.

\section{Model Structures and Data Preprocessing} \label{sec:model structures}
In this section we detail three different model structures which we subsequently compare on our experimental data:
\begin{enumerate}[A:]
    \item A simple model with a linear valve curve assumption.
    \item Parameterized, nonlinear valve curves.
    \item The same valve curves as B, but including data preprocessing for hysteresis compensation.
\end{enumerate}

\subsection{Model A: Naive Valve Curve Parameterization}
Our first and most simple model equates to the model considered in \cite{AGNER2023hydraulic_parameter_estimation}. Here we assume that all valves are fully linear and thus can be parameterized by only one function $k(\cdot)$:
\begin{equation}
    k(v_i) = v_i.
    \label{eq:linear valve curve}
\end{equation}

\subsection{Model B: Enhanced Valve Curve Parameterization}
\begin{figure}
    \centering
    \includegraphics[width=.7\textwidth]{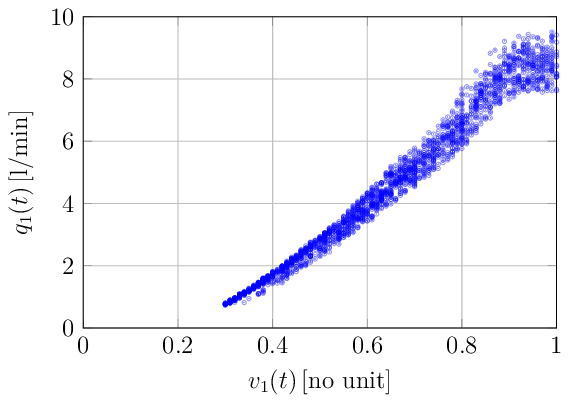}
    \caption{Scatter-plot of flow rates $q_1(t)$ against valve set-points $v_1(t)$ for valve 1 in the experiment setup to be described in the subsequent section.}
    \label{fig:scatter of q1 vs v1}
\end{figure}
To account for nonlinear valve characteristics, we consider also a larger number of parameterization functions $k_k(\cdot)$. To make an educated guess about reasonable valve curve shapes, \cref{fig:scatter of q1 vs v1} shows a scatter plot of $q_1(t)$ over $v_1(t)$ for valve 1 in the experimental setup to be described later. While this scatter-plot does not account for changes in differential pressure over the valve, we can clearly see that a linear curve from 0 to 1 does not capture this shape. Instead, it appears that the curve should have a sub-linear curve corresponding more to equal percentage characteristics. Additionally, the valve is clearly fully open already around $v_1 \approx 0.9$, and fully closed around $v_1 \approx 0.2$ To handle these nonlinear patterns, we consider valve characteristic functions on the form
\begin{equation}
    k_k(v_i) = \ramp{v_i}{a_k}{b_k}^{c_k}.
\end{equation}
for different choices of $a_k$, $b_k$ and $c_k$. Based on our intuition from \cref{fig:scatter of q1 vs v1}, we considered all such combinations corresponding to
\begin{align}
    a_k &\in \left[ 0.10, 0.15, 0.20, 0.25\right], \\
    b_k &\in \left[ 0.80, 0.85, 0.90, 0.95, 1.0\right], \\
    c_k &\in \left[ 1.0, 1.25, 1.5\right],
\end{align}
resulting in $K=60$ different valve parameterization functions $k_k(\cdot)$ for each valve. We chose the values for $a_k$ and $b_k$ as heuristically reasonable values for full opening and closing, and the values of $c_k$ as values corresponding to reasonable curves for equal percentage valves, which was deemed appropriate based on \cref{fig:scatter of q1 vs v1}.

\subsection{Model C: Hysteresis Compensation}
Our third and most sophisticated model uses the same valve characterization as model B. Model C however accounts for the effect of hysteresis. This is an effect which typically manifests in control valves, causing an issue where the valve spindle does not move for small changes in the set-point given to the valve \cite[p. 88]{tore_practice}. We account for this issue in the following way. Instead of using the valve set-points $v_i(t)$ for model fitting and validation, we utilize filtered versions $\hat{v}_i(t)$ which are meant to estimate the actual valve spindle position. We first define a hand-tuned parameter $\delta = 0.015$. The value $0.015$ was found heuristically by testing the values $0.05, 0.10, 0.15, 0.20$ and then choosing the best one. The spindle is assumed to remain still whenever the set-point moves a distance less than $\delta$ from the current estimated spindle position. Once the reference moves outside of this range, the estimated spindle position will lag behind by a distance $\delta$. These assumptions are captured in the following filter:
\begin{equation}
    \hat{v}_i(t) = \begin{cases}
        \hat{v}_i(t-1) & \text{if } \left| v_i(t) - \hat{v}_i(t-1)\right| \leq \delta \\
        v_i(t) - \delta & \text{if } v_i(t) \geq \hat{v}_i(t-1) + \delta \\
        v_i(t) + \delta & \text{else.}
    \end{cases}
\end{equation}
The filtered time series are initiated with the first measured valve set-point $\hat{v}_i(0) = v_i(0)$.

\section{Experiment Description} \label{sec:experiment}

Our experiment setup represents four houses, connected to a line-structured district heating network supplied by a single production facility. This is the same structure that we used as an example in Section \ref{sec:background} as seen in \cref{fig:graph-example}. 

% \begin{figure}[h!]
%     \centering
%     \resizebox{.5\linewidth}{!}{
%     \input{fig/hydraulic-schematic}
%     }
%     \caption{The hydraulic scheme of our experiment setup with 4 houses along a line-structured district heating grid, which corresponds to a representation of the network presented in \cref{fig:graph-example} \todo{We should perhaps include measurements here}.}
%     \label{fig:hydraulic experiment setup}
% \end{figure}

\subsection{Equipment}

The experiments were conducted in Aalborg University's Smart Water Infrastructure Lab (SWIL). SWIL is a state-of-the-art water and district heating laboratory that consists of configurable modules which can be used to build a wide variety of experimental setups  \cite{smart_water_lab}. The available module types are:

\begin{itemize}
    \item Pumping modules 
    \item Consumer modules
    \item Piping modules
    \item Heating modules
\end{itemize}

\begin{figure}[h!]
    \centering
    \includegraphics[width=\textwidth]{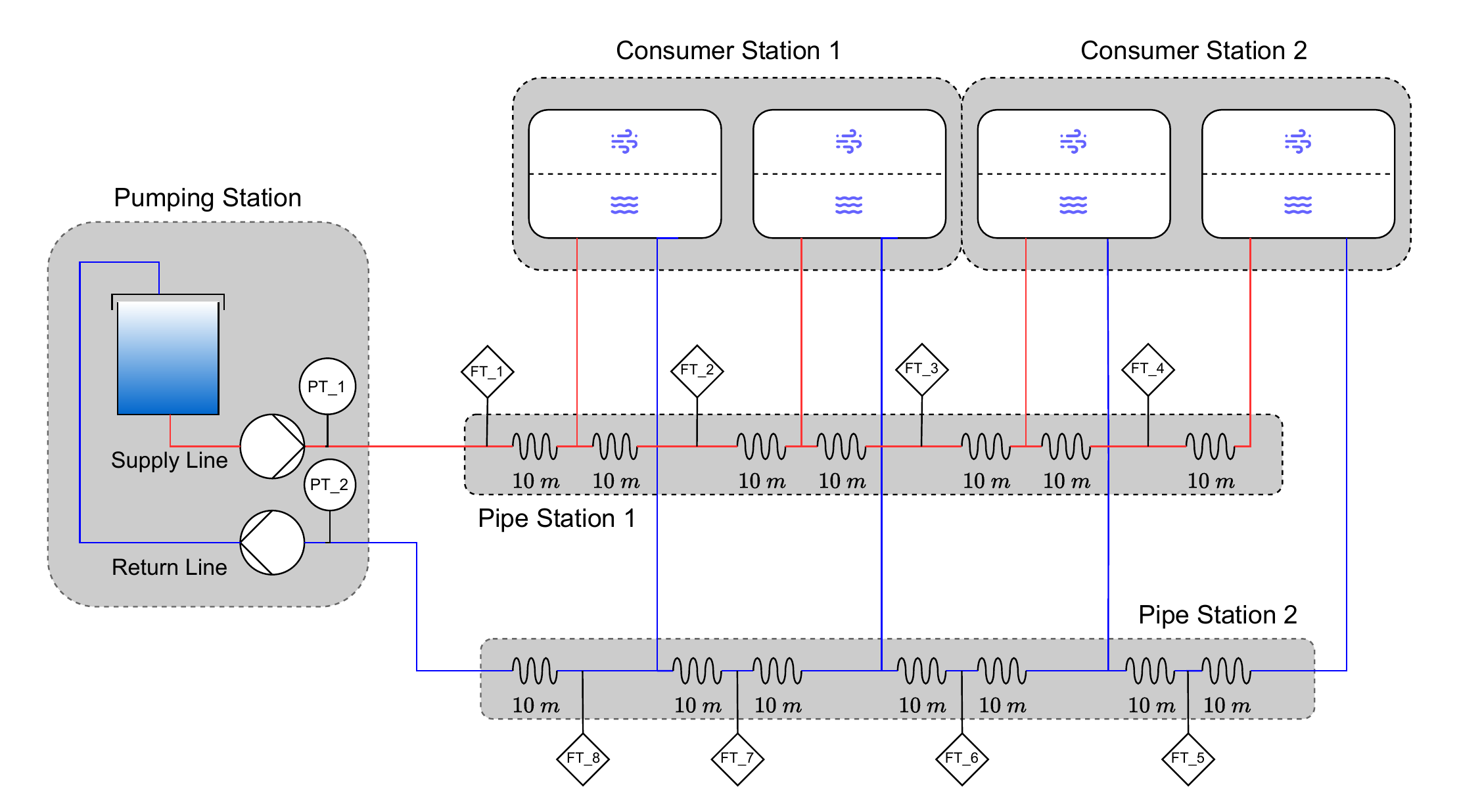}
    \caption{Diagram of the experimental setup used to recreate \cref{fig:graph-example}. \texttt{PT} and \texttt{FT} denote respectively flow and pressure transmitters, while the two blocks in each consumer unit are water-to-air heat exchangers.}
    \label{fig:AgnerLab}
\end{figure}

\begin{figure}
    \centering
    \begin{subfigure}[t]{.48\textwidth}
        \centering
        \includegraphics[width=.7\textwidth]{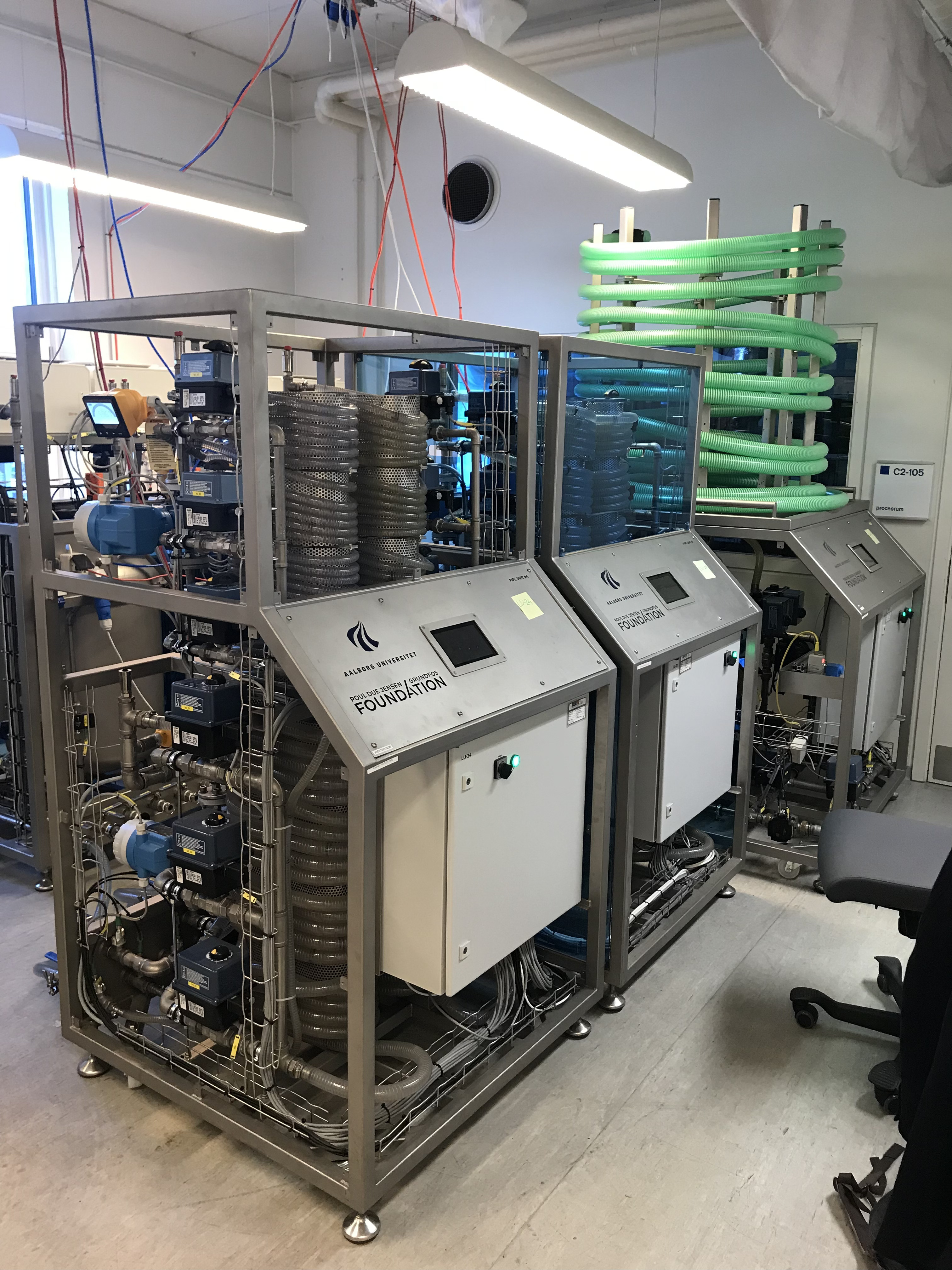}
        \caption{On the left are two pipe units. These units measure volume flow rates and pressures, and can be configured for pipe sections of varying lengths. On the right is a pumping station, where water is pumped from a water reservoir into the system, until finally returning to the reservoir.}
        \label{fig:lab pipes and pump}
    \end{subfigure}
    \hfill
    \begin{subfigure}[t]{.48\textwidth}
        \centering
        \includegraphics[width=.7\textwidth]{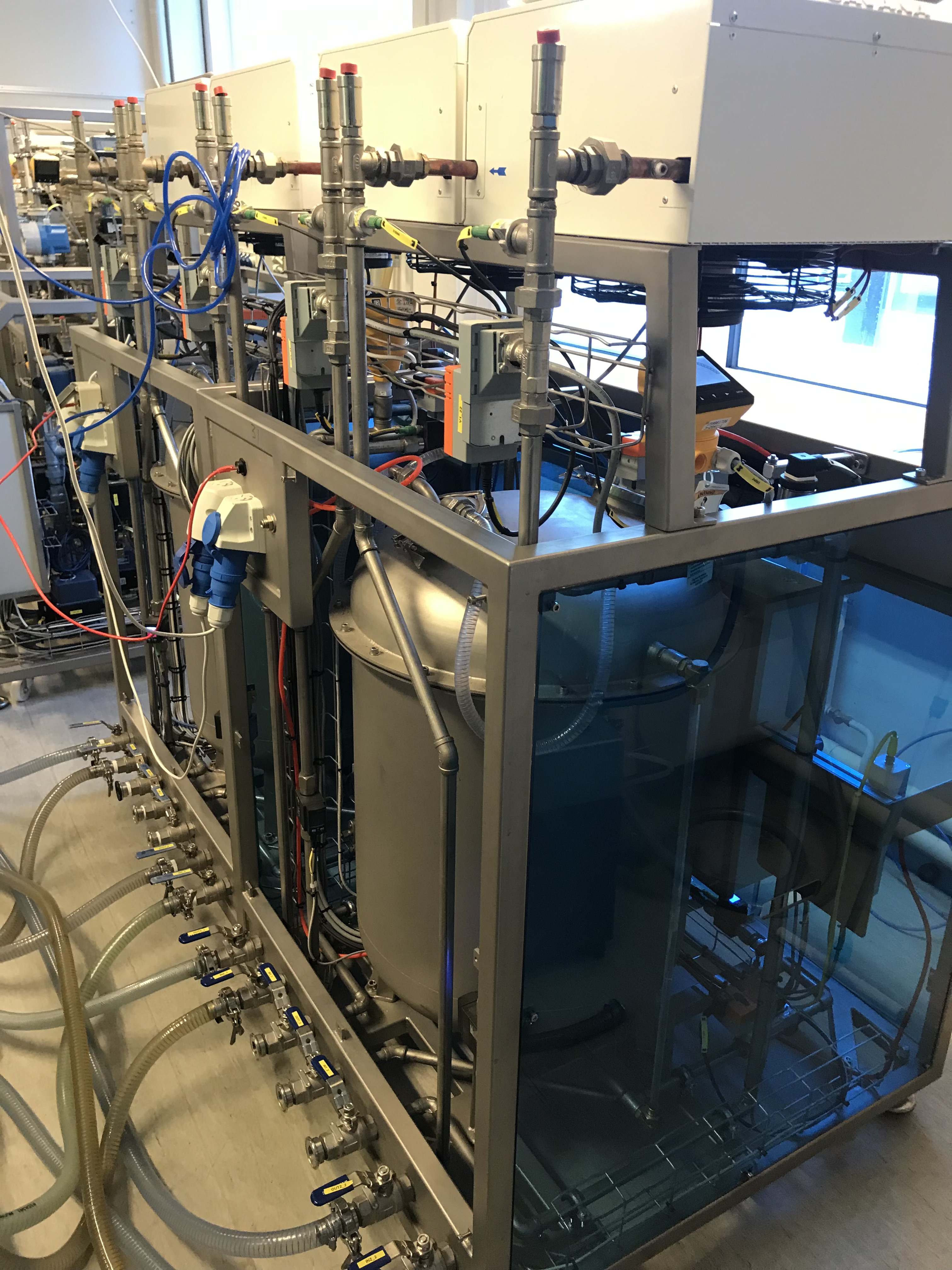}
        \caption{Two consumer units, which in total simulate four consumers. The bottom sections are connected to the pipe units. Water then flows up and through the water-to-air heat exchangers at the top, via a control valve.}
        \label{fig:lab consumers}
    \end{subfigure}
    \caption{Images from the laboratory setup.}
    \label{fig:lab-equipment}
\end{figure}

Examples of consumer and piping modules are seen in \cref{fig:lab-equipment}. Pumping and consumer modules are particularly configurable and may be used to emulate e.g. gravity sewers or elevated reservoirs. The experimental network corresponding to the schematic \cref{fig:graph-example} consists of a pumping station, two piping stations, and two consumer stations as seen on \cref{fig:AgnerLab}

\subsection{Measurements}
Flows and pressure were measured using the piping module sensors, respectively an Endress \& Hauser Proline Promag 10 and Grundfos Direct Sensor RPI+T $0-1.6$, connected to Beckhoff I/O modules. The data was collected at $1$ Hz and transmitted from the laboratory modules to a central control unit (CCU). The laboratory modules are controlled by soft PLCs comprising CodeSys software running on a Raspberry Pi. The CCU communicates with the modules over Modbus TCP through a Simulink simulation using the MATLAB Industrial Communications Toolbox and the Simulink Real-Time Pacer \cite{Vallabha2016}, and raw measurements are converted to respectively \qunit and \punit at this stage. The consumer volume flow rates $q_1, \dots, q_4$ were subsequently calculated as
\begin{align}
    q_1 &= q_\texttt{FT\_1} - q_\texttt{FT\_2}, \\
    q_2 &= q_\texttt{FT\_2} - q_\texttt{FT\_3}, \\
    q_3 &= q_\texttt{FT\_3} - q_\texttt{FT\_4}, \\
    q_4 &= q_\texttt{FT\_4}
\end{align}
where e.g. $q_\texttt{FT\_1}$ is the volume flow rate measured in sensor \texttt{FT\_1}. The pressure measurements \texttt{PT\_1} and \texttt{PT\_2} were used to represent the network root pressures $p_\alpha$ and $p_\beta$ respectively.

\subsection{Data Sets}
We generated two experimental datasets, henceforth referred to as the \textit{exciting} and \textit{realistic} data sets. 

\begin{figure}
    \centering
    \input{fig/step-response-data}
    \caption{Illustration of the method used to generate the \textit{exciting} data set. Valve set-points are randomly generated every 40 seconds. The subsequent 10 seconds of data is discarded to remove transient effects, and the mean of the remaining 30 seconds of data is logged as a data point. Valve set-points are lower-bounded by 0.3.}
    \label{fig:step illustration}
\end{figure}
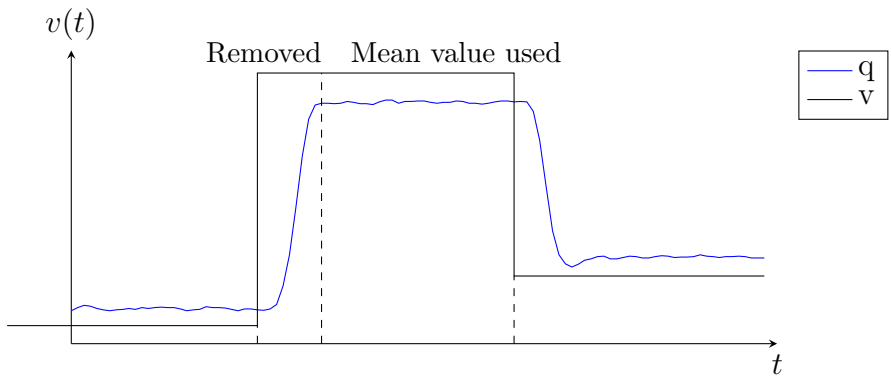

In the \textit{exciting} data set, we kept the pump working at full capacity and set the valve set-points randomly to individually and uniformly sampled vales in the interval $\left[0.3, 1.0\right]$ every 40 seconds. The lower bound of $0.3$ corresponds to valves which are in practice almost fully closed. This bound was included to avoid that a large part of the data corresponds to closed valves. During the full experiment, the pump was operating constantly at full capacity. In post-processing, the first 10 seconds in each 40 second interval was removed in order to ignore the transient phase of each step. The mean of remaining 30 seconds was used to generate a data point for later model fitting and validation. This process is shown in \cref{fig:step illustration}.

\begin{figure}
    \centering
    \input{fig/realistic-data}
    \caption{ Flow rates $q_i(t)$ in the realistic data set, recorded during the experiment. 48 minutes corresponds to 24 hours of real world time, in which we can see typical daily consumption patterns. }
    \label{fig:realistic data}
\end{figure}
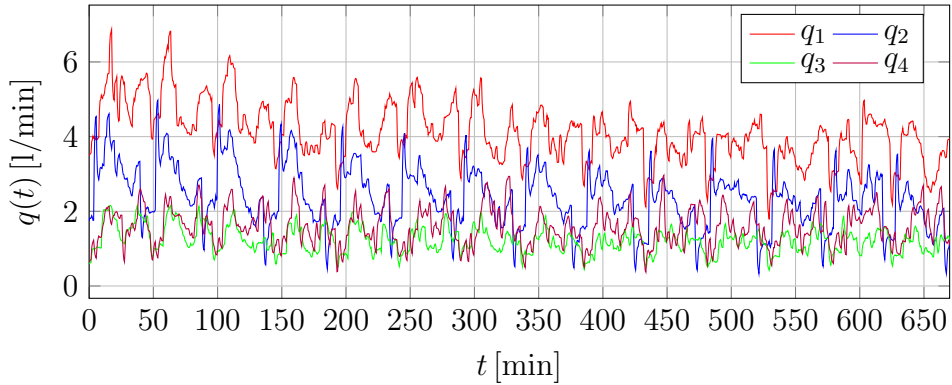

The exciting data set was produced to give us the best possible chance of accurately estimating the system parameters, but would be hard to replicate in a real district heating network. In the \textit{realistic} data set, the valves were instead configured to regulate flow rates which emulate consumption patterns from a real district heating system. These resulting volume flow rates are shown in \cref{fig:realistic data}. The valves were controlled with hand-tuned PID-controllers tracking the flow rate through each valve. The reference values for these flow rates were re-scaled time-series recorded in a district heating system in Nuremberg during 2022. The reference flow rates correspond to two weeks of hourly mean flow rates. While the valves regulated the system flow rates, the pump was operating constantly at full capacity to ensure sufficient system pressurization. For the purpose of experiment tractability, we scaled down the time such that 1 hour of reference data corresponded to 2 minutes in the experiment. We used mean values of 30 seconds of data for the subsequent model fitting and validation which corresponds to 15 minute mean value measurements in the real district heating setting. The reference volume flow rates correspond to the combined load for domestic hot water and space heating. In a real system, these loads are often separated into two different hydraulic loops controlled by separate valves \cite[p.365]{thebible}, but such a separation would be infeasible in our small laboratory setup.

\subsection{Model Evaluation}
We choose to evaluate the predictive power of the model from a given set of control inputs (valve set-points $v_i$ and differential pressure $\Delta p_0$ provided by the pump) to system outputs (flow rates though the four control valves representing consumers). Appendix A describes how to use the parameterized models to calculate estimated volume flow rates $\hat{q}_i$ in this way. We then investigate the prediction errors $e_i = q_i - \hat{q}_i$ to judge the model performance. This is a metric which is relevant when using the model to design flow rate controllers. We can note that this is not the objective for which the parameters are tuned. The training objective \eqref{eq:optimal cost} rather reflects the ability of the model to predict pressure drops given the valve set-points and flow rates in the system. The component-wise modeling of the network is what allows this flexibility in model application. This multi-purpose capacity is testimony to the value of considering grey-box component-models of this type over black-box models such as neural networks or regression models without structure considerations. There are other relevant choices of evaluation that could also be considered. In \cite{AGNER2023hydraulic_parameter_estimation}, ground-truth parameter value comparison was used. This is however clearly infeasible in the real setting where no ground truth is available.

\section{Results} \label{sec:results}
The following section details the results of the experiments. Firstly the model structures A, B and C trained on the exciting data set are investigated. Afterwards we evaluate the best performing model structure C trained on the realistic data set.

\subsection{Exciting training data}
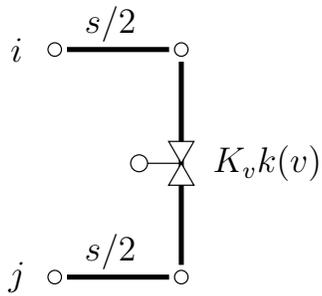
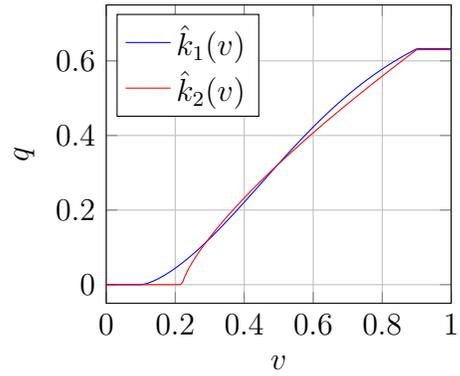
\begin{figure}
    \centering
    \begin{subfigure}[t]{.45\textwidth}
        \centering
        \resizebox{.8\textwidth}{!}{
        \input{fig/valve-pipe-overlap-hydraulics}
        }
        \vspace{.4cm}
        \caption{Hydraulic schematic of one consumer in the district heating grid, represented by two symmetrical pipes with combined resistance $s$ and valve curve $K_v k(v)$.}
        \label{fig:valve pipe schematic}
    \end{subfigure}
    \hfill
    \begin{subfigure}[t]{.45\textwidth}
        \centering
        \input{fig/valve_overlap_example}
        \caption{Plots of $\hat{k}_1(v)$ and $\hat{k}_2(v)$ which relate the flow rate $q$ and the valve position $v$ when $p_i-p_j = 1$.}
        \label{fig:valve pipe plots}
    \end{subfigure}
    \caption{An illustration of a consumer connected to the grid, represented by two symmetrical pipes and a valve. For a large portion of the valve curve, two different models of the hydraulic admittance of this consumer could have very similar properties. Here $\hat{k}_1(v)$ represents a pipe and a valve, but $\hat{k}_2(v)$ represents only a valve with no pipe. The difference between the two models becomes evident only when the valve is almost closed.}
    \label{fig:valve pipe comparison}
\end{figure}

\cref{tab:pipe parameters} shows the resulting pipe parameters $s_i$ using the different models. Pipes 1-4 lead to only one of the four consumers, whereas pipes 5-7 are shared between the consumers in the sense that the e.g. the volume flow rate for all four consumers comes through pipe 5. Of interest is that the resistances $s_1$, $s_2$ and $s_3$ were estimated as 0 for all three models trained on the exciting data set. One explanation for this is that our data does not represent the section of the valve curve where the valve is fully closed. To see why this matters, consider \cref{fig:valve pipe schematic}. Here we see a hydraulic schematic representing one consumer in the grid, i.e. two symmetric pipes (here represented by a total resistance $s$) and a valve (with curve $K_vk(v)$). The relationship between $\Delta p = p_i - p_j$, i.e. the pressure difference between supply-and-return-lines where the consumer is connected, and the flow rate $q$ for the consumer will be
\begin{equation}
    \Delta p = \left(s + \frac{1}{K_v^2 k(v)^2} \right)q^2
\end{equation}
and therefore
\begin{equation}
    q = \frac{K_v k(v)}{\sqrt{s K_v^2 k(v)^2} + 1}\sqrt{\Delta p}.
\end{equation}
We can denote $\hat{k}(v) = \frac{K_v k(v)}{\sqrt{s K_v^2 k(v)^2} + 1}$.  Consider two such choices of $\hat{k}(v)$: Firstly, $\hat{k}_1(v)$ where $s=1.5$, $k(v) = \ramp{v}{0.1}{0.9}^{1.5}$ and $K_v = 1$. Secondly, $\hat{k}_2(v)$ with $s=0$ (no pipe), $k(v) = \ramp{v}{0.22}{0.9}^{0.75}$, $K_v = 0.63$. \cref{fig:valve pipe plots} shows the plots of $\hat{k}_1(v)$ and $\hat{k}_2(v)$. Here we see that if we consider only valve-positions above $v = 0.3$, the curves are almost equivalent even though one of them has no pipe resistance. In the same way, due to the fact that we chose to omit valve openings below 0.3 in the training data, the model can not clearly discriminate between the valve curve parameters and the pipe parameters. This is an important realization, demonstrating that effective data sets should include also valve set-points in the lower range of operation. There are a couple of possible explanations as to why $s_4 \neq 0$ for models B and C while the other three are 0. Firstly pipes 1-3 should be equivalent due to the experiment configuration, with pipe 4 being longer than the others and thus more resistive. Secondly, since consumer 4 is the furthest from the pump, it also experiences the lowest differential pressure, which causes the volume flow rate to occasionally drop to 0. This could influence the estimation procedure towards $s_4 \neq 0$. Pipes 5, 6 and 7 do not experience the same issue as discussed with regards to \cref{fig:valve pipe comparison}, most likely because they are not in strict series with a valve. Pipes 6 and 7 should be similar in size. However, in models B and C, pipe 7 appears roughly a factor 5 longer than pipe 6, indicating that either the model does not capture the contribution of the pipes fully, or there are other reasons such as small twists and turns in the piping causing a difference between pipes 5 and 6. 

\begin{table}[]
    \centering
    \newcommand{\zero}{\textcolor{darkgray}{0.0}}
    \begin{tabular}{c|c||c|c|c|c||c|c|c}
        Model & Data & $s_1$ & $s_2$ & $s_3$ & $s_4$ & $s_5$ & $s_6$ & $s_7$ \\ \hline
        A & E & \zero & \zero & \zero & \zero & 0.0089 & 0.00082 & 0.021 \\
        B & E & \zero & \zero & \zero & 0.00067 & 0.0039 & 0.0046 & 0.029 \\
        C & E & \zero & \zero & \zero & 0.015 & 0.0038 & 0.0045 & 0.029 \\
        C & R & \zero & \zero & \zero & \zero & 0.0044 & \zero & 0.049  
    \end{tabular}
    \caption{Pipe parameters estimated for each model structure on either the exciting (E) or realistic (R) dataset.}
    \label{tab:pipe parameters}
\end{table}

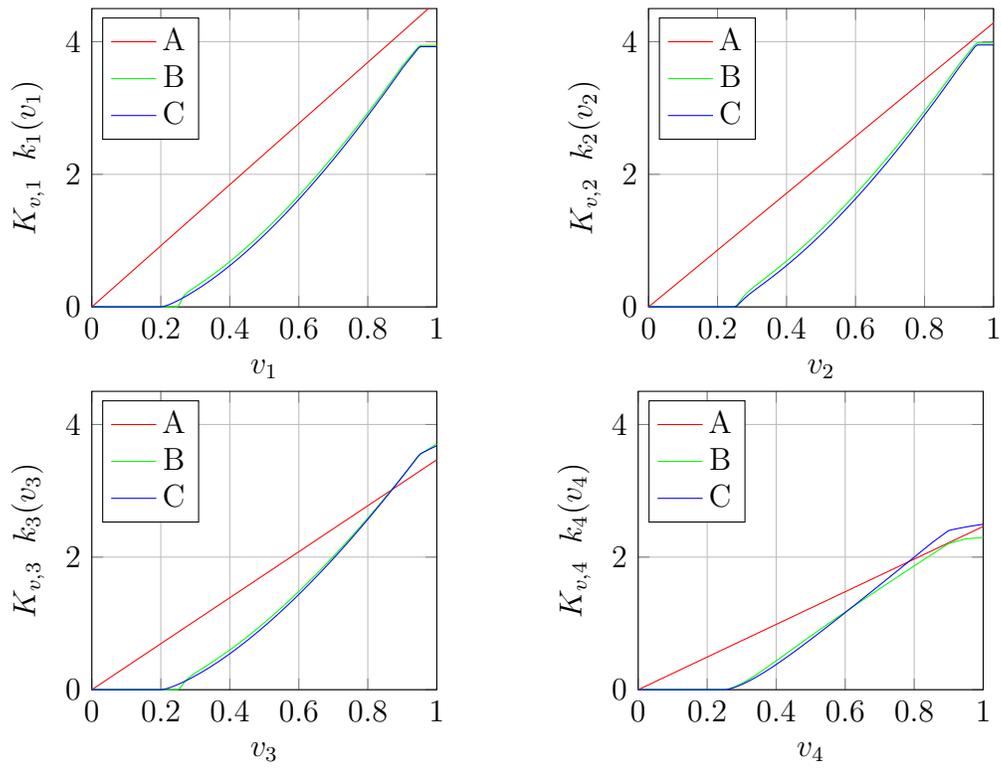
\begin{figure}
    \centering
    \input{fig/results/exciting_valve_curves}
    \caption{Valve curves $K_{v, i} k_i(v_i)$ for each valve $i$ under each model structure A, B and C trained on the exciting data set.}
    \label{fig:exciting valve curves}
\end{figure}

The resulting valve curve models become very sparse, an effect likely caused by the positivity constraint \eqref{eq:constraint theta positive} which prevents overfitting. Even though there are in total $K=60$ parameters $\theta_{i,k}$ for each valve $i$, resulting in 240 total valve curve parameters, only one of the valve curves has more than 5 non-zero parameters $\theta_{i,k}$. In total, model B has 18 parameters $\theta_{i,k} > 0$ and model C has 16. This model sparsity motivates the decision to omit parameter regularization from the scope of this paper. The full equation for each valve curve and each model as per \eqref{eq:valve characteristics} is provided in Appendix \ref{app: valve equations}. We can also investigate the valve characteristics visually, which yields \cref{fig:exciting valve curves}. As all four consumers are constructed with equal equipment, we would expect the four valve characteristic curves to be equivalent. However, this is not the case. Rather, valves 1 and 2 have similar characteristics, whereas valves 3 and 4 have much lower curves. This means that given equal differential pressure and valve set-points, the model predicts that valve 4 would yield a lower volume flow rate than the other valves. One possible explanation for this behavior is that the exponent $\gamma$ in \eqref{eq:pipe base model} and $q_i^2$ in \eqref{eq:valve characteristics} do not perfectly capture the real behavior of the system. If these exponents are too high, it means that the expected pressure drops at high flow rates will be over-estimated. Clearly, the volume flow rates and valve openings are strongly correlated (see \cref{fig:scatter of q1 vs v1}), hence the model can counter-act this effect by increasing the admittance at high valve set-point values. The reason this could make valve 4 stand out from the others is that the average volume flow rates over valve 4 are much lower than than the other valves (the mean volume flow rates are $\Bar{q}_1 = 4.60$ \qunit, $\Bar{q}_2 = 4.34$ \qunit, $\Bar{q}_3 = 2.91$ \qunit, and $\Bar{q}_4 = 2.11$ \qunit respectively) and thus the model does not need to over-estimate the admittance at high flow rates to the same extent.

% \begin{figure}
%     \centering
%     \input{fig/results/exciting_scatterplots}
%     \caption{Prediction errors $e_i = q_i - \hat{q}_i$ plotted over the valve set-point $v_i$ for each valve $i$ and each model structure A, B and C trained on the exciting data set. The plots are arranged as valve 1-4 from top to bottom and model A (red), B (green), and C (blue) from left to right.}
%     \label{fig:scatter exciting}
% \end{figure}
\begin{figure}
    \centering
    \begin{subfigure}[t]{0.32\textwidth}
        \includegraphics[width=\textwidth]{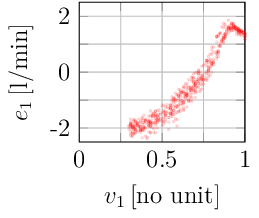}
    \end{subfigure}\hfill
    \begin{subfigure}[t]{0.32\textwidth}
        \includegraphics[width=\textwidth]{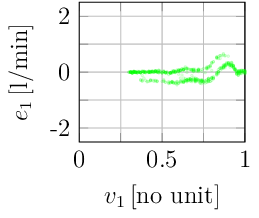}
    \end{subfigure}\hfill
    \begin{subfigure}[t]{0.32\textwidth}
        \includegraphics[width=\textwidth]{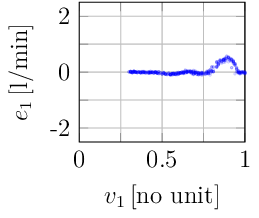}
    \end{subfigure}
    
    \begin{subfigure}[t]{0.32\textwidth}
        \includegraphics[width=\textwidth]{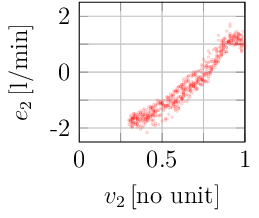}
    \end{subfigure}\hfill
    \begin{subfigure}[t]{0.32\textwidth}
        \includegraphics[width=\textwidth]{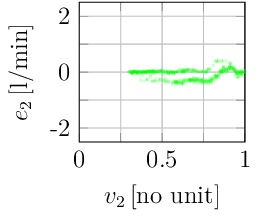}
    \end{subfigure}\hfill
    \begin{subfigure}[t]{0.32\textwidth}
        \includegraphics[width=\textwidth]{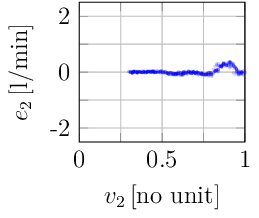}
    \end{subfigure}
    
    \begin{subfigure}[t]{0.32\textwidth}
        \includegraphics[width=\textwidth]{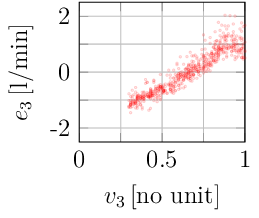}
    \end{subfigure}\hfill
    \begin{subfigure}[t]{0.32\textwidth}
        \includegraphics[width=\textwidth]{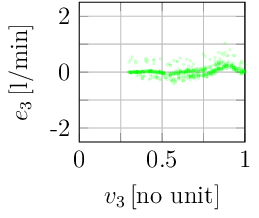}
    \end{subfigure}\hfill
    \begin{subfigure}[t]{0.32\textwidth}
        \includegraphics[width=\textwidth]{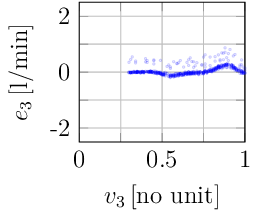}
    \end{subfigure}
    
    \begin{subfigure}[t]{0.32\textwidth}
        \includegraphics[width=\textwidth]{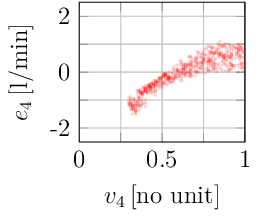}
    \end{subfigure}\hfill
    \begin{subfigure}[t]{0.32\textwidth}
        \includegraphics[width=\textwidth]{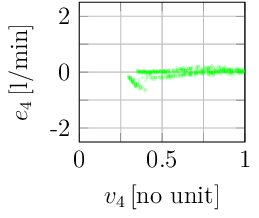}
    \end{subfigure}\hfill
    \begin{subfigure}[t]{0.32\textwidth}
        \includegraphics[width=\textwidth]{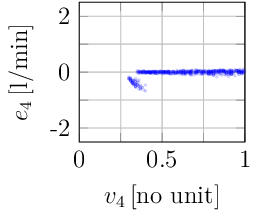}
    \end{subfigure}
    \caption{Prediction errors $e_i = q_i - \hat{q}_i$ plotted over the valve set-point $v_i$ for each valve $i$ and each model structure A, B and C trained on the exciting data set. The plots are arranged as valve 1-4 from top to bottom and model A (red), B (green), and C (blue) from left to right.}
    \label{fig:scatter exciting}
\end{figure}

\begin{figure}
    \centering
    \includegraphics[width=.65\textwidth]{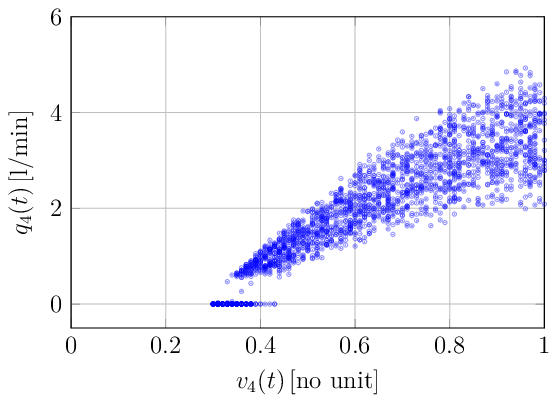}
    \caption{Scatter-plot of flow rates $q_4(t)$ against valve set-points $v_4(t)$ in the exciting data set.}
    \label{fig:scatter of q4 vs v4}
\end{figure}

\cref{fig:scatter exciting} shows the prediction errors $e_i(t) = q_i(t) - \hat{q}_i(t)$ plotted over the corresponding valve set-points $v_i(t)$. Each column represents one model structure A, B or C, and each row represents one valve index $i = 1,\dots,4$. \cref{fig:scatter exciting} shows that A has a strong correlation between prediction errors and valve set-points, which means that this naive parameterization of the valve curves is clearly insufficient. Both models B and C have a much lower correlation between valve set-points and prediction errors, with a few exceptions. Valves 1-3 display strong correlation between $e_i$ and $v_i$ around $v_i \approx 0.85$ which is the inflection point where the valve enters the fully closed position. It is possible that further tuning of the chosen parameterization $k_k(\cdot)$ could remove this issue. Valve 4 shows a cluster of outlier errors around $v_4 \approx 0.3$. We can find an explanation in \cref{fig:scatter of q4 vs v4} where we see the flow rate $q_4(t)$ scattered over the valve set-point $v_4(t)$ in the exciting data set. We can note that there are barely any recorded volume flow rates between 0 and 0.3 \qunit. This means that when the valve is almost closed and the differential pressure over the valve is low, there is a nonlinear behavior where the volume flow rate sharply cuts from 0.3 to 0, which is clearly not captured in the model. Valve 4 is the valve which is the furthest from the pump and hence experiences the lowest differential pressure. Hence the volume flow rate in the other valves never becomes sufficiently low to exhibit this behavior. Finally there are outliers in the prediction errors $e_3$. These outliers are likely caused by measurement errors which can occur due to e.g. air bubbles forming in the system. 

\begin{figure}
    \centering
    \begin{subfigure}[t]{0.42\textwidth}
        \includegraphics[width=\textwidth]{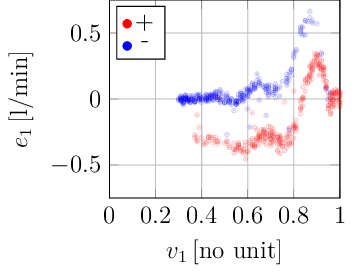}
    \end{subfigure}\hfill 
    \begin{subfigure}[t]{0.42\textwidth}
        \includegraphics[width=\textwidth]{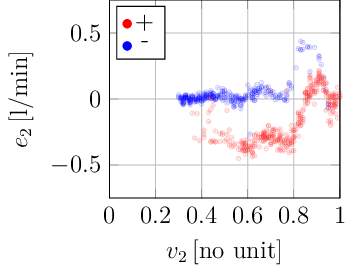}
    \end{subfigure}

    \begin{subfigure}[t]{0.42\textwidth}
        \includegraphics[width=\textwidth]{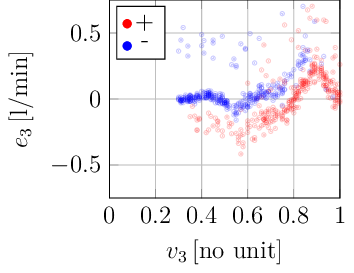}
    \end{subfigure}\hfill 
    \begin{subfigure}[t]{0.42\textwidth}
        \includegraphics[width=\textwidth]{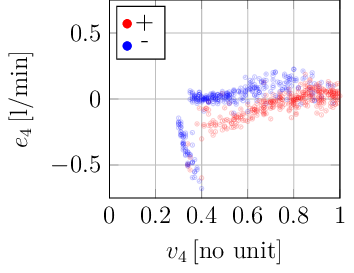}
    \end{subfigure}
    \caption{Prediction errors $e_i = q_i - \hat{q}_i$ for each valve $i$ using model B trained on the exciting data set, scattered over the valve set-points $v_i$. The points where the valve opened since the last data point ($v_i(t) - v_i(t-1) > 0$) are shown in red, and the points where the valve closed are shown in blue.}
    \label{fig:hysteresis plot}
\end{figure}

One difference between models B and C is that the prediction errors for model B appear segmented in one upper and one lower cluster. This is most clearly visible in valves 1 and 2. In \cref{fig:hysteresis plot} we show an enlarged view of these plots. Here we highlight the data points where the change in valve set-point is positive in red ($v_i(t) - v_i(t-1) > 0$), and negative in blue. The two clusters are clearly separated by the movement direction of the valve set-point, indicating the effect of hysteresis. In model C, this effect is no longer visible, demonstrating the strength of the hysteresis compensation.

% \begin{figure}
%     \centering
%     \input{fig/results/hysteresis_plot}
%     \caption{Prediction errors $e_i = q_i - \hat{q}_i$ for each valve $i$ using model B trained on the exciting data set, scattered over the valve set-points $v_i$. The points where the valve opened since the last data point ($v_i(t) - v_i(t-1) > 0$) are shown in red, and the points where the valve closed are shown in blue.}
%     \label{fig:hysteresis plot}
% \end{figure}

\begin{figure}
    \centering
    \begin{subfigure}[t]{0.42\textwidth}
        \includegraphics[width=\textwidth]{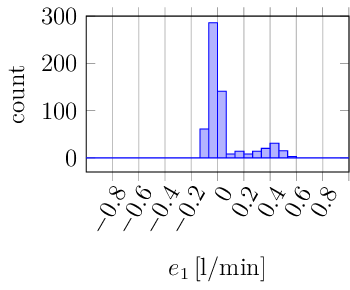}
    \end{subfigure}\hfill 
    \begin{subfigure}[t]{0.42\textwidth}
        \includegraphics[width=\textwidth]{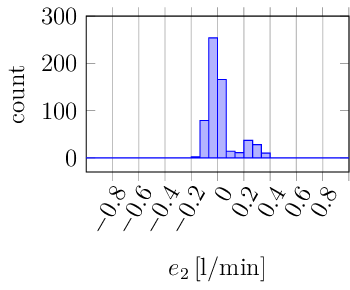}
    \end{subfigure}

    \begin{subfigure}[t]{0.42\textwidth}
        \includegraphics[width=\textwidth]{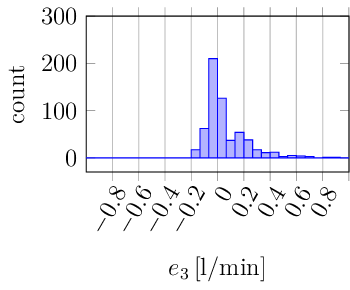}
    \end{subfigure}\hfill 
    \begin{subfigure}[t]{0.42\textwidth}
        \includegraphics[width=\textwidth]{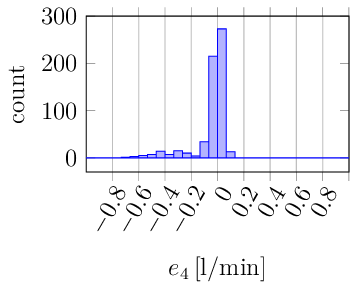}
    \end{subfigure}
    \caption{Histograms of prediction errors $e_i = q_i - \hat{q}_i$ using model structure C trained on the exciting data set.}
    \label{fig:c hist exciting}
\end{figure}

In total, model C is the strongest model when trained and tested on the exciting data set. Histograms of the prediction errors $e_i(t)$ for this model are shown in \cref{fig:c hist exciting}. All four error distributions have large tails. In \cref{fig:scatter exciting} we saw that these tails arise due to not fully capturing the valve characteristics for valves 1 and 2 and the data outliers of valves 3 and 4. Outside of these tails, a vast majority of the errors lie in the range $\left[-0.2,0.2\right] $\qunit, which can be compared with the mean flow rates $\Bar{q}_i$ exhibited in the system, given by $\Bar{q}_1 = 4.60$ \qunit, $\Bar{q}_2 = 4.34$ \qunit, $\Bar{q}_3 = 2.91$ \qunit, and $\Bar{q}_4 = 2.11$ \qunit respectively in the exciting data set. Hence in the well-tuned regions of the valve curve, errors remain within 5 or 10 \% of the mean flow rates for valves 1-2 and 3-4 respectively.

% \begin{figure}
%     \centering
%     \input{fig/results/c_hist_exciting}
%     \caption{Histograms of prediction errors $e_i = q_i - \hat{q}_i$ using model structure C trained on the exciting data set.}
%     \label{fig:c hist exciting}
% \end{figure}

\subsection{Realistic training data}

\begin{figure}
    \centering
    \input{fig/results/valve_curves_exciting_vs_realistic}
    \caption{Valve curves $K_{v,i} k_i(v_i)$ for each valve $i$ for model structure C trained on the exciting (E, red) and realistic (R, blue) datasets respectively.}
    \label{fig:valves exciting vs realistic}
\end{figure}
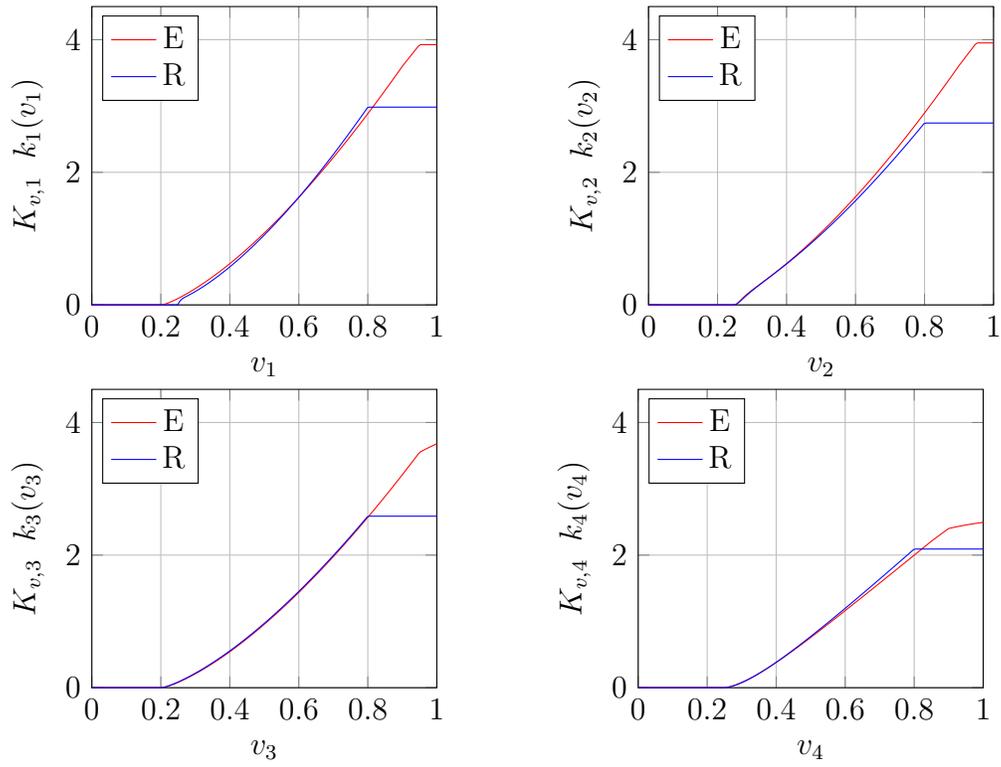

\begin{figure}
    \centering
    \begin{subfigure}[t]{0.42\textwidth}
        \includegraphics[width=\textwidth]{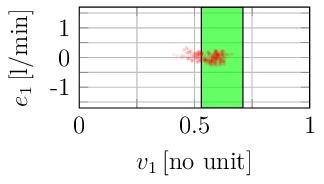}
    \end{subfigure}\hfill 
    \begin{subfigure}[t]{0.42\textwidth}
        \includegraphics[width=\textwidth]{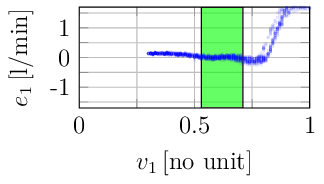}
    \end{subfigure}

    \begin{subfigure}[t]{0.42\textwidth}
        \includegraphics[width=\textwidth]{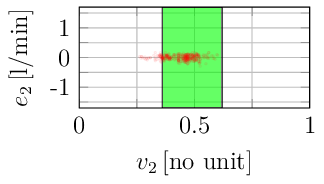}
    \end{subfigure}\hfill 
    \begin{subfigure}[t]{0.42\textwidth}
        \includegraphics[width=\textwidth]{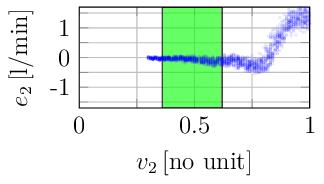}
    \end{subfigure}

    \begin{subfigure}[t]{0.42\textwidth}
        \includegraphics[width=\textwidth]{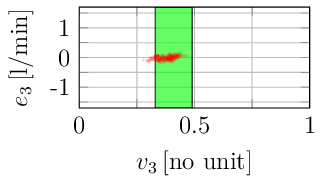}
    \end{subfigure}\hfill 
    \begin{subfigure}[t]{0.42\textwidth}
        \includegraphics[width=\textwidth]{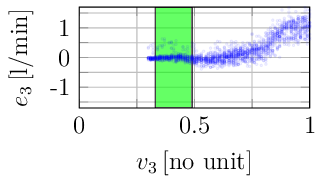}
    \end{subfigure}

    \begin{subfigure}[t]{0.42\textwidth}
        \includegraphics[width=\textwidth]{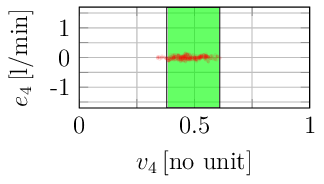}
    \end{subfigure}\hfill 
    \begin{subfigure}[t]{0.42\textwidth}
        \includegraphics[width=\textwidth]{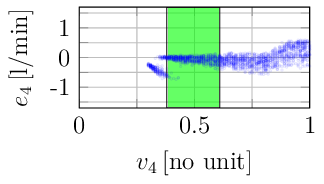}
    \end{subfigure}
    \caption{Prediction errors $e_i = q_i - \hat{q}_i$ for each valve $i$ using model C trained on the realistic data set and evaluated on both the realistic data set (red, left) and the exciting data set (blue, right). The figures are arranged as valve 1 to 4 from top to bottom. The green bar outlines the interval between the 5th and 95th quantiles for $v_i$ in the realistic training data.}
    \label{fig:scatter exciting vs realistic}
\end{figure}

\cref{fig:valves exciting vs realistic} shows the evaluated valve curves gained from training model structure C on the exciting or realistic data set respectively. We can note that in the lower regions of the valve curves, there is a significant overlap between the two curves. However, above a valve set-point of roughly 0.6, we see the curves starting to deviate. This is likely due to the fact that the model trained on the realistic data set has not seen any data within this region. This is also reflected in \cref{fig:scatter exciting vs realistic} which shows the prediction errors of using model C trained on the realistic data set. The model behaves rather well when evaluated on the realistic data set and maintains the errors within about 0.3 \qunit. However, the realistic data only represents valve set-points in limited intervals. The green bar in the figures highlights the interval between the 5th and 95th quantiles of the valve set-points in the training data. The model is unable to extrapolate beyond the limited intervals of valve set-points represented in the training data which is reflected in the plots on the right of \cref{fig:scatter exciting vs realistic}. Here the model is evaluated on the exciting data set, which covers a larger range of valve operation. In the region of the valve set-points where the model was trained, prediction errors are relatively small. Outside of these intervals however, the models behave extremely poorly. This is not surprising, as no such data was used in training. While it may be difficult to guarantee that the full operational range of each valve is represented in data from a real system, it may also be unlikely that the operational range present in the data is quite so restrictive as in our example. Firstly, if such a small range of a valves operational ranges is used, this would imply that the valve may be poorly dimensioned for the building in which it is installed. Secondly, we are only using two weeks worth of representational data. We can see already in \cref{fig:scatter exciting vs realistic} that the last 30 \% of the data has been shifted in comparison to the training data. Hence if a longer period of time is considered, it becomes increasingly likely that a larger portion of the valves operational range is explored. Thirdly, the reference values for this data are based on hourly means of recorded volume flow rates. The mean operation naturally hides the peaks and valleys of the operational modes and hence one could expect larger variance in a real data set.

% \begin{figure}
%     \centering
%     \input{fig/results/scatter_exciting_vs_realistic}
%     \caption{Prediction errors $e_i = q_i - \hat{q}_i$ for each valve $i$ using model C trained on the realistic data set and evaluated on both the realistic data set (red, left) and the exciting data set (blue, right). The figures are arranged as valve 1 to 4 from top to bottom. The green bar outlines the interval between the 5th and 95th quantiles for $v_i$ in the realistic training data.}
%     \label{fig:scatter exciting vs realistic}
% \end{figure}

\section{Conclusion} \label{sec:conclusion}
In this paper we modeled the hydraulic properties of a district heating system, and tuned the parameters of said model based on operational data. This method was evaluated on laboratory data from the Smart Water Infrastructure laboratory in Aalborg. We found that in a realistic setting, the characteristics of valves are hugely important for accurately describing the system properties. A simple assumption of linear valve characteristics is therefore a naive approach. To remedy this issue, we proposed modeling the valve characteristic with a linear combination of many possible valve characteristics, which improved model performance. Additionaly, we found that valve hysteresis played a significant role in the setup, which we remedied through the use of a preprocessing filter before doing model calibration. We compared the modeling procedure on two data set, one designed to test the system in all of its' possible operational mode and one designed to mimic an operational district heating system. We found that in the more realistic data set, the valves only operated in a limited region. This led to the model extrapolating poorly to operational modes outside of this region. However, in the first data set where a larger portion of the systems operational range was explored, the best performing model was able to predict volume flow rates within 5 and 10 \% deviation from the mean values, excluding the portion of the valve curve where the fit was the most poor. The performance in this region could likely be improved by further tailoring the parameterization of the valve curve.

\section{Future work} \label{sec:future work}
There are several open questions and directions for extending the results of this paper. Firstly, it is not common in current district heating systems to have access to measurements of valve positions. While this data may become available with future smart installations, a current approach could be to jointly estimate the model parameters and the valve positions. Such an approach would present many interesting challenges, as the quality of the parameter estimation depends on the accuracy of the valve position estimation, and the valve position estimation conversely depends on the accuracy of the model calibration.

A further investigation can be conducted with regards to choosing the filtering parameter $\delta$ for hysteresis compensation. In this experimental setup we have only four valves which can be assumed to have similar characteristics. Thus a shared, heuristically chosen $\delta$ is viable. For a larger and more heterogeneous network, this approach is no longer valid and thus a reasonable search approach is needed.

It appears that a further investigation into the exponent $\gamma$ relating volume flow rates and differential pressures over pipes and valves could be an avenue to model improvement. However, such an investigation can quickly become complex. E.g. should this exponent be shared between valves and pipes, and should each component have its' own exponent? Tuning this parameter based on data is a non-trivial task with regards to maintaining computational tractability.

Finally one can consider an extension to meshed distribution networks. This causes the issue that the volume flow rates in pipes in the supply-and-return networks are not explicitly known by measuring the valve volume flow rates. Previous works have solved for these values in tandem with model parameters using non-convex optimizers such as particle swarm solvers \cite{zheng_thermo-hydraulic_2023}. Such previous works have however been limited to considering only pipe resistances and not valve characteristics. Therefore they could be less concerned with the computational complexity and convergence issues which may differentiate non-convex from convex problems. 

\section*{Acknowledgements}
Felix Agner and Anders Rantzer are members of the ELLIIT Strategic Research Area at Lund University. 
% Christian M. Jensen and Carsten S. Kallesøe are with Grundfos Holdings A/S and Aalborg University, Department of Electronic Systems, Automation
% \& Control Section (e-mail: \{chjensen,ckallesoe\}@grundfos.com). Rafal Wisniewski is with Aalborg University, Department of Electronic Systems, Automation \& Control Section (e-mail: raf@es.aau.dk).

This work is funded by the European Research Council (ERC) under the European Union's Horizon 2020 research and innovation program under grant agreement No 834142 (ScalableControl). 

The work has been partially supported by the Poul Due Jensen Foundation through the Smart Water Infrastructure Laboratory project and the Swift project. 

Christian Møller Jensen is funded by Danish Innovation Fund grant no. 3129-00019B.

\section*{Data Availability}
The data sets prepared and used in this paper are openly available via GitHub under a CC-BY 4.0 license \cite{data_v0}.
%% If you have bibdatabase file and want bibtex to generate the
%% bibitems, please use
%%
\bibliographystyle{elsarticle-num} 
\bibliography{bibliography}

%% else use the following coding to input the bibitems directly in the
%% TeX file.

% \begin{thebibliography}{00}

% %% \bibitem{label}
% %% Text of bibliographic item

% \bibitem{}

% \end{thebibliography}
\appendix
\section*{Appendix A - Forward estimation} \label{app: forward estimation}
\counterwithin{equation}{section}  % Reset the equation counter for each section
\renewcommand{\theequation}{A.\arabic{equation}}  % Change the equation label format
\renewcommand{\thefigure}{A.\arabic{figure}}
\setcounter{figure}{0}

\begin{figure}
    \centering
    \begin{subfigure}{.4\textwidth}
        \centering
        \resizebox{\textwidth}{!}{
        \input{fig/forward-estimation/full_network}
        }
        \caption{Valve resistance $r_4$ and pipe resistances $s_4$ are replaced by one equivalent resistance $\hat{s}_4$.}
        \label{fig:forward full}
    \end{subfigure} \hfill
    \begin{subfigure}{.33\textwidth}
        \centering
        \resizebox{\textwidth}{!}{
        \input{fig/forward-estimation/reduced_by_one}
        }
        \caption{The parallel connection of valves 3 and 4 are reduced to one equivalent resistance $\hat{s}_3$.}
        \label{fig:forward reduced one}
    \end{subfigure} \hfill
    \begin{subfigure}{.2\textwidth}
        \centering
        \resizebox{.45\textwidth}{!}{
        \input{fig/forward-estimation/fully_reduced}
        }
        \caption{The network is reduced to one equivalent resistance $\hat{s}_1$.}
        \label{fig:forward fully reduced}
    \end{subfigure}
    \caption{Step-by-step illustration for reducing the full hydraulic representation of the network into one equivalent resistance with given valve set-points.}
    \label{fig:forward estimation}
\end{figure}
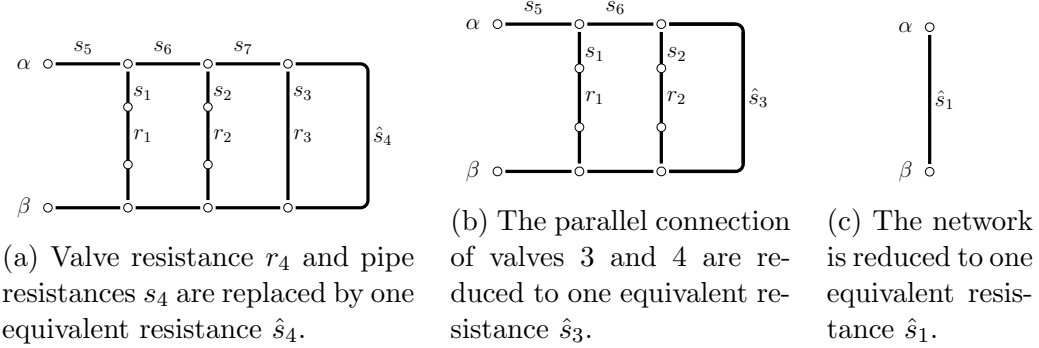

In a line-structured grid such as the one we investigate, our parameterized model can be used for forward estimation. This implies calculating the resulting flow rates $q_i$ given the valve set-points $v_i$ and the differential pressure $\Delta p_0$. To accomplish this we can realize that given a parameterized valve model on the form \eqref{eq:valve model} and given a valve set-point $v_i$,  \eqref{eq:valve model} can be replaced by
\begin{equation}
    \Delta p_i = r_i q_i^2
\end{equation}
where $r_i = \left(\frac{\theta_{i,1}}{k_1(v_i)^2} + \frac{\theta_{i,2}}{k_2(v_i)^2} + \dots + \frac{\theta_{i,K}}{k_K(v_i)^2} \right)$. In this way once the spindle position is fixed, a valve functions like a pipe in the sense of how it relates pressure drops and flow rates. To calculate the total amount of water flowing from $\alpha$ to $\beta$, we can reduce our system in a series of steps to one single, equivalent resistance (see \cref{fig:forward estimation}). First we replace $s_4$ and $r_4$ with an equivalent resistance $\hat{s}_4 = 2s_4 + r_4$ (see \cref{fig:forward full}). We can now find a relation between $q_3$ and $q_4$ by establishing that 
$$ \hat{s}_4 q_4^2 = (2s_3 + r_3)q_3^2$$
and hence
$$ q_3 = \sqrt{\frac{\hat{s}_4}{2s_3 + r_3}}q_4.$$
We will now lump the part of the network right of valve 2 into one equivalent resistance $\hat{s}_3$ (see \cref{fig:forward reduced one}). The flow rate going through $\hat{s}_3$ is $q_3+q_4$, and hence it must hold that
\begin{align}
    \hat{s}_3(q_3+q_4)^2 & = 2s_6\left(q_3+q_4\right)^2 + \hat{s}_4q_4^2 \\
    \implies \hat{s}_3 & = 2s_6 + \frac{q_4^2}{(q_3+q_4)^2}\hat{s_4} \nonumber \\
    & = 2s_6 + \frac{q_4^2}{q_4^2\left(\sqrt{\hat{s}_4/(2s_3 + r_3)} + 1\right)^2}\hat{s}_4 \nonumber \\
    & = 2s_6 + \frac{\left(2s_3 + r_3\right)\hat{s}_4}{\sqrt{\hat{s}_4} + \sqrt{2s_3 + r_3}}.
\end{align}
We can iteratively calculate $\hat{s}_i$ in this way according to
\begin{equation}
    \hat{s}_i = 2s_{i+3} + \frac{\left(2s_i + r_i\right)\hat{s}_{i+1}}{\sqrt{\hat{s}_{i+1}} + \sqrt{2s_i + r_i}}
    \label{eq:iterative shat}
\end{equation}
until reaching $\hat{s}_1$ which directly connects $\alpha$ to $\beta$ (see \cref{fig:forward fully reduced}). Now clearly the total flow rate $\hat{q}_1 = q_1+q_2+q_3+q_4$ must satisfy
\begin{equation}
    \hat{q}_1 = \sqrt{\frac{\Delta p_0}{\hat{s}_1}}.
\end{equation}
Now we can calculate the individual flow rates $q_1$, $q_2$, $q_3$ and $q_4$ by expanding the network again in the other direction. We here note that it must hold that
\begin{align*}
    \hat{s}_1\hat{q}_1^2 & = \hat{s}_2\left(\hat{q}_1 - q_1\right)^2 \\
    \implies q_1 & = \left(1 - \sqrt{\frac{\hat{s}_1}{\hat{s}_2}}\right)\hat{q}_1
\end{align*}
and also that $\hat{q}_2 = q_2+q_3+q_4 = \hat{q}_1-q_1$. We can thus calculate all of the flow rates recursively according to the rules
\begin{align}
    q_i &= \left(1 - \sqrt{\frac{\hat{s}_i}{\hat{s}_{i+1}}}\right)\hat{q}_i \\
    \hat{q}_{i+1} &= \hat{q}_i - q_i.
\end{align}

\section*{Appendix B - Valve Characteristic Equations} \label{app: valve equations}
\counterwithin{equation}{section} 
\renewcommand{\theequation}{B.\arabic{equation}} 
\renewcommand{\thefigure}{B.\arabic{figure}}
\setcounter{figure}{0}
\setcounter{equation}{0}
The expressions relating flow rates, valve set-points and differential pressure for the valves in each model can be written down on closed form. For model A trained on the exciting data set, this relation is given by
\begin{subequations}
    \begin{align}
        \Delta p_1 &= \frac{ 0.047 }{v_1^2}q_1^2 \\ \Delta p_2 &= \frac{ 0.054 }{v_2^2}q_2^2 \\ \Delta p_3 &= \frac{ 0.083 }{v_3^2}q_3^2 \\ \Delta p_4 &= \frac{ 0.16 }{v_4^2}q_4^2.
    \end{align}
\end{subequations}
For model B trained on the exciting data set, the expression is given by
\begin{subequations}
    \begin{align}
        \Delta p_1 &= \scalebox{0.985}{$\left(\frac{ 0.011 }{\ramp{ v_1 }{ 0.15 }{ 0.9 }^{ 3.0 } } +\frac{ 0.049 }{\ramp{ v_1 }{ 0.15 }{ 0.95 }^{ 3.0 } } +\frac{ 0.0041 }{\ramp{ v_1 }{ 0.2 }{ 0.95 }^{ 3.0 } } +\frac{ 0.00015 }{\ramp{ v_1 }{ 0.25 }{ 0.95 }^{ 3.0 } }\right)$}q_1^2 \\ 
        \Delta p_2 &= \scalebox{0.80}{$\left(\frac{ 0.009 }{\ramp{ v_2 }{ 0.1 }{ 0.9 }^{ 3.0 } } +\frac{ 0.0068 }{\ramp{ v_2 }{ 0.1 }{ 0.95 }^{ 3.0 } } +\frac{ 0.039 }{\ramp{ v_2 }{ 0.15 }{ 0.95 }^{ 3.0 } } +\frac{ 0.0068 }{\ramp{ v_2 }{ 0.2 }{ 0.95 }^{ 3.0 } } +\frac{ 0.0011 }{\ramp{ v_2 }{ 0.25 }{ 0.95 }^{ 3.0 } }\right)$}q_2^2 \\ 
        \Delta p_3 &= \scalebox{0.985}{$\left(\frac{ 0.039 }{\ramp{ v_3 }{ 0.15 }{ 0.95 }^{ 3.0 } } +\frac{ 0.029 }{\ramp{ v_3 }{ 0.15 }{ 1.0 }^{ 3.0 } } +\frac{ 0.0044 }{\ramp{ v_3 }{ 0.2 }{ 1.0 }^{ 3.0 } } +\frac{ 0.00072 }{\ramp{ v_3 }{ 0.25 }{ 1.0 }^{ 3.0 } }\right)$}q_3^2 \\ 
        \Delta p_4 &= \scalebox{0.81}{$\left(\frac{ 0.11 }{\ramp{ v_4 }{ 0.1 }{ 0.9 }^{ 2.0 } } +\frac{ 0.031 }{\ramp{ v_4 }{ 0.1 }{ 1.0 }^{ 2.0 } } +\frac{ 0.0043 }{\ramp{ v_4 }{ 0.2 }{ 0.9 }^{ 3.0 } } +\frac{ 0.0078 }{\ramp{ v_4 }{ 0.25 }{ 0.9 }^{ 3.0 } } +\frac{ 0.033 }{\ramp{ v_4 }{ 0.25 }{ 0.95 }^{ 3.0 } }\right)$}q_4^2 
    \end{align}
\end{subequations}
For model C trained on the exciting data set, the expressions are
\begin{subequations}
    \begin{align} 
        \Delta p_1 &= \scalebox{1.0}{$ \left(\frac{ 0.0093 }{\ramp{ v_1 }{ 0.15 }{ 0.9 }^{ 3.0 } } +\frac{ 0.03 }{\ramp{ v_1 }{ 0.15 }{ 0.95 }^{ 3.0 } } +\frac{ 0.025 }{\ramp{ v_1 }{ 0.2 }{ 0.95 }^{ 3.0 } }\right) $}q_1^2 \\ 
        \Delta p_2 &= \scalebox{0.985}{$ \left(\frac{ 0.0065 }{\ramp{ v_2 }{ 0.15 }{ 0.9 }^{ 3.0 } } +\frac{ 0.038 }{\ramp{ v_2 }{ 0.15 }{ 0.95 }^{ 3.0 } } +\frac{ 0.018 }{\ramp{ v_2 }{ 0.2 }{ 0.95 }^{ 3.0 } } +\frac{ 0.0022 }{\ramp{ v_2 }{ 0.25 }{ 0.95 }^{ 3.0 } }\right)$}q_2^2 \\ 
        \Delta p_3 &= \scalebox{1.0}{$ \left(\frac{ 0.04 }{\ramp{ v_3 }{ 0.15 }{ 0.95 }^{ 3.0 } } +\frac{ 0.0089 }{\ramp{ v_3 }{ 0.2 }{ 0.95 }^{ 3.0 } } +\frac{ 0.025 }{\ramp{ v_3 }{ 0.2 }{ 1.0 }^{ 3.0 } }\right)$}q_3^2 \\ 
        \Delta p_4 &= \scalebox{0.66}{$ \left(\frac{ 0.023 }{\ramp{ v_4 }{ 0.1 }{ 0.85 }^{ 2.0 } } +\frac{ 0.069 }{\ramp{ v_4 }{ 0.1 }{ 0.9 }^{ 2.0 } } +\frac{ 0.0085 }{\ramp{ v_4 }{ 0.2 }{ 0.9 }^{ 3.0 } } +\frac{ 0.036 }{\ramp{ v_4 }{ 0.25 }{ 0.9 }^{ 3.0 } } +\frac{ 0.001 }{\ramp{ v_4 }{ 0.25 }{ 0.95 }^{ 3.0 } } +\frac{ 0.023 }{\ramp{ v_4 }{ 0.25 }{ 1.0 }^{ 3.0 } }\right)$}q_4^2 
    \end{align} 
\end{subequations}
Finally, when trained on the realistic data set, model C yields the following expressions.
\begin{subequations}
    \begin{align}
        \Delta p_1 &= \left(\frac{ 0.11 }{\ramp{ v_1 }{ 0.2 }{ 0.8 }^{ 3.0 } } +\frac{ 0.00014 }{\ramp{ v_1 }{ 0.25 }{ 0.8 }^{ 3.0 } }\right)q_1^2 \\ 
        \Delta p_2 &= \left(\frac{ 0.11 }{\ramp{ v_2 }{ 0.15 }{ 0.8 }^{ 3.0 } } +\frac{ 0.024 }{\ramp{ v_2 }{ 0.25 }{ 0.8 }^{ 2.5 } } +\frac{ 0.0029 }{\ramp{ v_2 }{ 0.25 }{ 0.8 }^{ 3.0 } }\right)q_2^2 \\ 
        \Delta p_3 &= \left(\frac{ 0.085 }{\ramp{ v_3 }{ 0.15 }{ 0.8 }^{ 3.0 } } +\frac{ 0.065 }{\ramp{ v_3 }{ 0.2 }{ 0.8 }^{ 3.0 } }\right)q_3^2 \\ 
        \Delta p_4 &= \left(\frac{ 0.077 }{\ramp{ v_4 }{ 0.1 }{ 0.8 }^{ 2.0 } } +\frac{ 0.026 }{\ramp{ v_4 }{ 0.1 }{ 0.8 }^{ 2.5 } } +\frac{ 0.12 }{\ramp{ v_4 }{ 0.25 }{ 0.8 }^{ 3.0 } }\right)q_4^2.
    \end{align}
\end{subequations}

%% file: fig/grid-schematic.tex
\begin{tikzpicture}

    % layers
    \pgfdeclarelayer{supplynodes}
    \pgfdeclarelayer{returnnodes}
    \pgfdeclarelayer{supplyedges}
    \pgfdeclarelayer{returnedges}
    \pgfdeclarelayer{valves}
    \pgfdeclarelayer{buildings}
    \pgfdeclarelayer{annotation}
    \pgfdeclarelayer{highlight}
    
    \pgfsetlayers{highlight, returnedges, valves, returnnodes, supplyedges, supplynodes, buildings, annotation, main}

    % plotting parameters
    \def\xOff{2.7cm} % offset in x between nodes
    \def\xOffHouse{1.1cm} % offset in x between main line and houses
    \def\xOffPlant{3.5cm} % offset to the plant
    \def\yOff{2.4cm} % offset in y between nodes
    \def\yOffHouse{1.5cm}
    \def\nRad{0.8mm} % radius of nodes
    \def\edgeWidth{1.2mm} % width of edges
    % node colors
    \def\rColor{white}
    \def\sColor{black}
    %path highlight
    \def\hlColor{green}
    \def\hlWidth{3.5mm}
    
    \begin{pgfonlayer}{returnnodes}
        % draw base nodes beta, 1 and 5
        \node[circle, fill=blue, minimum size=\nRad, text=\rColor, font=\boldmath] (beta) at (0,0) {$\mathbf{\beta}$};
        \node[circle, fill=blue, minimum size=\nRad, text=\rColor] (r5) at (\xOffPlant,0) {$\mathbf{5}$};
        \node[circle, fill=blue, minimum size=\nRad, text=\rColor, xshift=\xOffHouse, yshift=\yOffHouse] (r1) at (r5) {$\mathbf{1}$};

        % draw all the other nodes
        \foreach \i/\ilast in {6/5, 7/6, 2/1, 3/2, 4/3}{
        \node[circle, fill=blue, minimum size=\nRad, text=\rColor, xshift=\xOff] (r\i) at (r\ilast) {$\mathbf{\i}$};
        }
    \end{pgfonlayer}

    \begin{pgfonlayer}{supplynodes}
        % draw alpha
        \node[circle, fill=red, minimum size=\nRad, text=\rColor, yshift=\yOff, font=\boldmath] (alpha) at (beta) {$\mathbf{\alpha}$};
        % draw all the other nodes
        \foreach \i in {1,...,7}{
        \node[circle, fill=red, minimum size=\nRad, text=\rColor, yshift=\yOff] (s\i) at (r\i) {$\mathbf{\i}$};
        }
    \end{pgfonlayer}

    \begin{pgfonlayer}{returnedges}
        % do special one for 7-4
        \coordinate[xshift=\xOff] (rCorner) at (r7); 
        \draw[blue, line width=\edgeWidth, rounded corners=10pt] (r7) to node[currarrow, rotate=180] {} (rCorner) to node[currarrow, rotate=240] {} (r4);
        % add all the other edges
        \foreach \i/\j\r in {beta/r5/180, r5/r6/180, r6/r7/180, r1/r5/240, r6/r2/240, r7/r3/240}{
        \draw[blue, line width=\edgeWidth] (\i) to node[currarrow,rotate=\r] {} (\j);
        }
    \end{pgfonlayer}

    \begin{pgfonlayer}{supplyedges}
        % do special one for 7-4
        \coordinate[xshift=\xOff] (sCorner) at (s7); 
        \draw[red, line width=\edgeWidth, rounded corners=10pt] (s7) to node[currarrow] {} (sCorner) to node[currarrow, rotate=60] {} (s4);
        % add all the other edges
        \foreach \i/\j\r in {alpha/s5/0, s5/s6/0, s6/s7/0, s1/s5/60, s6/s2/60, s7/s3/60}{
        \draw[red, line width=\edgeWidth] (\i) to node[currarrow,rotate=\r] {} (\j);
        }
    \end{pgfonlayer}

    \begin{pgfonlayer}{valves}
        \foreach \i in {1,...,4}{
        \draw[black, line width=\edgeWidth] (s\i) to node[currarrow,rotate=270] {} (r\i);
        }
    \end{pgfonlayer}

    \begin{pgfonlayer}{buildings}
        % houses above connections

        % roofs for houses
        \foreach \i in {1,...,4}{
            \node[fill=black, minimum size=.5cm, text=white] (h\i) [above=of s\i, yshift=-.9cm] {};
            \draw[fill=black] ([yshift=-0.1mm]h\i.north) -- ++(-.3cm,0) -- ++(0.3cm,0.15cm) -- ++(0.3cm,-0.15cm) -- cycle;
        }
    
        % heat plant above root
        % define position
        \node (HP) [above=of alpha, yshift=-.9cm] {};
        % draw the shape
        \draw[fill=black] (HP) 
        -- ++(-.5cm,0) 
        -- ++ (0,.8cm) 
        -- ++ (.2cm,0) 
        -- ++ (0,-.4cm) 
        -- ++ (.25cm,.2cm) -- ++ (0,-.2cm) % jagged shape
        -- ++ (.25cm,.2cm) -- ++ (0,-.2cm) % jagged shape
        -- ++ (.25cm,.2cm) -- ++ (0,-.2cm) % jagged shape
        -- ++ (0,-.4cm)
        -- cycle;
    \end{pgfonlayer}

    \begin{pgfonlayer}{annotation}
        % valves and pipes 1-4
        \foreach \i in {1,...,4}{
        \node[xshift=0.3cm,yshift=-0.9cm,font=\boldmath,color=black] at (s\i) {$\i$};
        \node[xshift=-0.7cm,yshift=-0.4cm,font=\boldmath,color=red] at (s\i) {$\i$};
        \node[xshift=-0.7cm,yshift=-0.4cm,font=\boldmath,color=blue] at (r\i) {$\i$};
        }
        % pipes 5-7
        \foreach \i in {5,...,7}{
        \node[xshift=-0.9cm,yshift=-0.3cm,font=\boldmath,color=red] at (s\i) {$\i$};
        \node[xshift=-0.9cm,yshift=-0.3cm,font=\boldmath,color=blue] at (r\i) {$\i$};
        }
        
    \end{pgfonlayer}

    \begin{pgfonlayer}{highlight}
        \draw[\hlColor, line width=\hlWidth, opacity=0.5] (alpha) -- (s5) -- (s6) -- (s2) -- (r2) -- (r6) -- (r5) -- (beta);
        \node[xshift=-1.2cm,yshift=0.4cm,color=black,font=\boldmath] at (s5) {$\mathcal{L}_2$};
    \end{pgfonlayer}
    
\end{tikzpicture}

\iffalse

\fi

%% file: fig/valve_curves.tex
\begin{tikzpicture}
\begin{axis}[
    axis lines = left,
    width = .6\textwidth,
    height = .6\textwidth,
    xlabel = \(v_i\),
    ylabel = {\(k_i(v_i)\)},
    legend pos = north west,
    legend style = {
    font=\small
    },
    legend cell align=left,
    grid=major,
    xmin=0,xmax=1.05,
    ymin=0,ymax=1.05,
]
%linear curve
\addplot [
    domain=0:1, 
    samples=100, 
    color=red,
]
{x};
\addlegendentry{$v$}
%equal percentage
\addplot [
    domain=0.0:0.1, 
    samples=2, 
    color=blue,
    forget plot,
]{0};
\addplot [
    domain=0.8:1.0, 
    samples=2, 
    color=blue,
    forget plot,
]{1};
\addplot [
    domain=0.1:0.8, 
    samples=100, 
    color=blue,
]
{(x-0.1)^2 / (0.7^2)};
\addlegendentry{\(\ramp{v}{0.1}{0.8}^{2}\)}

%quick opening
\addplot [
    domain=0.0:0.2, 
    samples=2, 
    color=green,
    forget plot,
]{0};
\addplot [
    domain=0.9:1.0, 
    samples=2, 
    color=green,
    forget plot,
]{1};
\addplot [
    domain=0.2:0.9, 
    samples=100, 
    color=green,
]
{(x-0.2)^0.5 / (0.7^0.5)};
\addlegendentry{\(\ramp{v}{0.2}{0.9}^{0.5}\)}

\end{axis}
\end{tikzpicture}

%% file: fig/step-response-data.tex
\begin{tikzpicture}
\begin{axis}[
    axis lines = left,
    width = .8\textwidth,
    height = .4\textwidth,
    xlabel = $t$,
    ylabel = {$v(t)$},
    ymax = 6.5,
    ymin = 0,
    xmin = 10,
    xmax = 120,
    ticks = none,
    xlabel style={at={(current axis.right of origin)},anchor=north},
    ylabel style={at={(current axis.above origin)},rotate=-90,anchor=south},
    clip=false,
    legend pos=outer north east,  % position the legend in the top-right corner
]
% Step response
\addplot[blue, restrict x to domain=10:118] table[x=time, y=q, col sep=comma]{data/step_response.csv};
\addlegendentry{q}  % add an entry to the legend for the q plot
\draw (0,0.4) -- (39,0.4) -- (39,6) -- (79,6) -- (79,1.5) -- (118,1.5);
\addlegendimage{black, line legend}
\addlegendentry{v}  % add an entry to the legend for the v plot
% Time stamps
\draw[style=dashed] (39,0) -- (39,0.4);
\draw[style=dashed] (49,0) -- (49,6);
\draw[style=dashed] (79,0) -- (79,1.5);
\node[above] at (axis cs:40,6) {\small{Removed}};
\node[above] at (axis cs:70,6) {\small{Mean value used}};

\end{axis}
\end{tikzpicture}

%% file: fig/realistic-data.tex
\begin{tikzpicture}
\begin{axis}[
    width = .95\textwidth,
    height = .4\textwidth,
    xlabel = {$t \left[\text{min}\right]$},
    ylabel = {$q(t) \left[ \text{\qunit} \right]$},
    legend pos=north east, 
    enlarge x limits=false, % scale automatically to x data
    legend columns=2,
    grid = both,
]
% Step response
\addplot[red] table[x=lab_time_minutes, y=q0, col sep=comma]{data/realistic_downsample.csv};
\addlegendentry{$q_1$}
\addplot[blue] table[x=lab_time_minutes, y=q1, col sep=comma]{data/realistic_downsample.csv};
\addlegendentry{$q_2$}
\addplot[green] table[x=lab_time_minutes, y=q2, col sep=comma]{data/realistic_downsample.csv};
\addlegendentry{$q_3$}
\addplot[purple] table[x=lab_time_minutes, y=q3, col sep=comma]{data/realistic_downsample.csv};
\addlegendentry{$q_4$}

\end{axis}
\end{tikzpicture}

%% file: fig/valve-pipe-overlap-hydraulics.tex
\begin{tikzpicture}

    % layers
    \pgfdeclarelayer{nodes}
    \pgfdeclarelayer{edges}
    \pgfdeclarelayer{valves}
    \pgfdeclarelayer{pump}
    \pgfdeclarelayer{annotation}
    \pgfsetlayers{edges,nodes,valves,pump,annotation,main}

    \def\xOff{1.5cm} % offset in x between nodes
    \def\yOff{2.7cm} % offset in y between nodes
    \def\nRad{0.8mm} % radius of nodes
    \def\edgeWidth{0.6mm} % width of edges
    \def\pumpRad{3.0mm} % radius of pump symbol
    \def\valveSize{3.0mm} % size of valve symbols 

    \begin{pgfonlayer}{nodes}
        % Return-side nodes
        \node[] (r0) at (0,0) {};
        \node[xshift=\xOff] (r1) at (r0) {};
        
        % Supply-side nodes
        \node[yshift=\yOff] (s0) at (r0) {};
        \node[xshift=\xOff] (s1) at (s0) {};

        % draw the nodes
        \foreach \n in {r0, r1, s0, s1}{
        \draw[fill=white] (\n) circle (\nRad);
        }

    \end{pgfonlayer}

    \begin{pgfonlayer}{edges}
        % connect all the supplies and returns
        \foreach \s/\r in {s1/r1}{
        \draw[black, line width=\edgeWidth] (\s) -- (\r);
        }
        \foreach \si/\sj/\ri/\rj in {s0/s1/r0/r1} {
        \draw[black, line width=\edgeWidth] (\ri) -- (\rj);
        \draw[black, line width=\edgeWidth] (\si) -- (\sj);
        }
    \end{pgfonlayer}

    \begin{pgfonlayer}{pump}
        % draw a circle
        %\draw[fill=white] ($(s0)!.5!(r0)$) circle (\pumpRad);
        % draw triangle in circle
        %\draw[] ($(s0)!.5!(r0) +(210:\pumpRad)$) -- ($(s0)!.5!(r0) +(90:\pumpRad)$) -- ($(s0)!.5!(r0) +(330:\pumpRad)$);
    \end{pgfonlayer}

    \begin{pgfonlayer}{valves}
        \foreach \r/\s in {r0/s0}{
        % draw triangles
        \draw[fill=white] ($(\r)!.5!(\s) + (\xOff,0)$) --
        ($(\r)!.5!(\s) + (\xOff,0) + (60:\valveSize)$) --
        ($(\r)!.5!(\s) + (\xOff,0) + (120:\valveSize)$) --
        ($(\r)!.5!(\s) + (\xOff,0) + (300:\valveSize)$) --
        ($(\r)!.5!(\s) + (\xOff,0) + (240:\valveSize)$) -- cycle;
        % draw actuator symbol
        \draw[]{} ($(\r)!.5!(\s) + (\xOff,0)$) -- ($(\r)!.5!(\s) + (\xOff,0) + (-.5cm,0)$);
        \draw[fill=white] ($(\r)!.5!(\s) + (\xOff,0) + (-.5cm,0)$) circle (1mm);
        }
        
    \end{pgfonlayer}

    \begin{pgfonlayer}{annotation}
        %\foreach \r/\s/\i in {r0/s0/2}{
        % draw valve indices
        \node[] at ($(r0)!.5!(s0) + (\xOff,0) + (10.2mm,0)$) {$K_v k(v)$};
        % draw customer pipe indices
        %\node[] at ($(\r)!.5!(\s) + (\xOff,0) + (2.4mm,8.0mm)$) {\i};
        %}
        %\foreach \s/\i in {s0/s}{
        % draw pipe numbers for larger pipes
        \node[] at ($(s0) + (6.6mm, 3.0mm)$) {$s/2$};
        \node[] at ($(r0) + (6.6mm, 3.0mm)$) {$s/2$};
        %}
        %\foreach \n/\i in {s0/i, r0/j}{
        \node[] at ($(s0) + (-4.5mm,0)$) {$i$};
        \node[] at ($(r0) + (-4.5mm,0)$) {$j$};
        %}
    \end{pgfonlayer}
    
\end{tikzpicture}

%% file: fig/valve_overlap_example.tex
\begin{tikzpicture}
\begin{axis}[
    width = \textwidth,
    height = .9\textwidth,
    xlabel={$v$},
    ylabel={$q$},
    grid=major,
    xmin = 0, xmax = 1,
    ymin = -0.05, ymax = 0.75,
    legend entries={$\hat{k}_1(v)$, $\hat{k}_2(v)$},
    legend pos=north west,
]
\addplot[blue] table [x=v, y=f2, col sep=comma] {data/valve_overlap_example.csv};
\addplot[red] table [x=v, y=f1, col sep=comma] {data/valve_overlap_example.csv};

\end{axis}
\end{tikzpicture}

%% file: fig/results/exciting_valve_curves.tex
\newcommand{\valvecurve}[3]{
\begin{subfigure}[t]{0.45\textwidth}
    \begin{tikzpicture}
    \begin{axis}[
        width=\textwidth,
        height=.9\textwidth,
        xmin=0, xmax=1,
        ymin = 0, ymax = 4.5,
        grid = both,
        legend pos=north west,
        xlabel = {$v_#2$},
        ylabel = {$K_{v,#2} \; \; k_#2(v_#2)$},
    ]
    \addplot[red] table[y=#1, x=v, col sep=comma] {data/results/valve_curve_A_exciting.csv};
    \addlegendentry{A}
    \addplot[green] table[y=#1, x=v, col sep=comma] {data/results/valve_curve_B_exciting.csv};
    \addlegendentry{B}
    \addplot[blue] table[y=#1, x=v, col sep=comma] {data/results/valve_curve_C_exciting.csv};
    \addlegendentry{C}
    \end{axis}
    \end{tikzpicture}
    %\caption{#3}
\end{subfigure}
}

\valvecurve{kv0}{1}{}
\hfill
\valvecurve{kv1}{2}{}

\valvecurve{kv2}{3}{}
\hfill
\valvecurve{kv3}{4}{}

%% file: fig/results/valve_curves_exciting_vs_realistic.tex
\newcommand{\valvecurve}[3]{
\begin{subfigure}[t]{0.45\textwidth}
    \begin{tikzpicture}
    \begin{axis}[
        width=\textwidth,
        height=.9\textwidth,
        xmin=0, xmax=1,
        ymin = 0, ymax = 4.5,
        grid = both,
        legend pos=north west,
        xlabel = {$v_#2$},
        ylabel = {$K_{v,#2} \; \; k_#2(v_#2)$},
    ]
    \addplot[red] table[y=#1, x=v, col sep=comma] {data/results/valve_curve_C_exciting.csv};
    \addlegendentry{E}
    \addplot[blue] table[y=#1, x=v, col sep=comma] {data/results/valve_curve_C_realistic.csv};
    \addlegendentry{R}
    \end{axis}
    \end{tikzpicture}
    %\caption{#3}
\end{subfigure}
}

\valvecurve{kv0}{1}{}
\hfill
\valvecurve{kv1}{2}{}

\valvecurve{kv2}{3}{}
\hfill
\valvecurve{kv3}{4}{}

%% file: fig/forward-estimation/full_network.tex
\begin{tikzpicture}

    % layers
    \pgfdeclarelayer{nodes}
    \pgfdeclarelayer{edges}
    \pgfdeclarelayer{annotation}
    \pgfsetlayers{edges,nodes,annotation,main}

    \def\xOff{1.5cm} % offset in x between nodes
    \def\yOff{2.7cm} % offset in y between nodes
    \def\nRad{0.8mm} % radius of nodes
    \def\edgeWidth{0.6mm} % width of edges
    \def\pumpRad{3.0mm} % radius of pump symbol
    \def\valveSize{3.0mm} % size of valve symbols 

    \begin{pgfonlayer}{nodes}
        % Return-side nodes
        \node[] (r0) at (0,0) {};
        \node[xshift=\xOff] (r1) at (r0) {};
        \node[xshift=\xOff] (r2) at (r1) {}; 
        \node[xshift=\xOff] (r3) at (r2) {}; 
        
        % Supply-side nodes
        \node[yshift=\yOff] (s0) at (r0) {};
        \node[xshift=\xOff] (s1) at (s0) {};
        \node[xshift=\xOff] (s2) at (s1) {};
        \node[xshift=\xOff] (s3) at (s2) {}; 

        % draw the nodes
        \foreach \n in {r0, r1, r2, r3, s0, s1, s2, s3}{
        \draw[fill=white] (\n) circle (\nRad);
        }
        % draw valve nodes
        \foreach \r/\s in {r0/s0, r1/s1}{
        \draw[fill=white] ($(\r)!.3!(\s) + (\xOff,0)$) circle (\nRad);
        \draw[fill=white] ($(\r)!.7!(\s) + (\xOff,0)$) circle (\nRad);
        }

    \end{pgfonlayer}

    \begin{pgfonlayer}{edges}
        % connect all the supplies and returns
        \foreach \s/\r in {s1/r1, s2/r2, s3/r3}{
        \draw[black, line width=\edgeWidth] (\s) -- (\r);
        }
        \foreach \si/\sj/\ri/\rj in {s0/s1/r0/r1, s1/s2/r1/r2, s2/s3/r2/r3} {
        \draw[black, line width=\edgeWidth] (\ri) -- (\rj);
        \draw[black, line width=\edgeWidth] (\si) -- (\sj);
        }
        \draw[black, line width=\edgeWidth] (s3) [rounded corners] -- ++(\xOff,0)  -- ($(r3) + (\xOff,0)$) -- (r3);
    \end{pgfonlayer}

    \begin{pgfonlayer}{annotation}
        \foreach \r/\s/\i in {r0/s0/1, r1/s1/2, r2/s2/3}{
        % draw valve indices
        \node[] at ($(\r)!.5!(\s) + (\xOff,0) + (2.8mm,0)$) {$r_\i$};
        % draw customer pipe indices
        \node[] at ($(\r)!.5!(\s) + (\xOff,0) + (2.8mm,8.0mm)$) {$s_\i$};
        }
        \foreach \s/\i in {s0/5, s1/6, s2/7}{
        % draw pipe numbers for larger pipes
        \node[] at ($(\s) + (6.6mm, 3.0mm)$) {$s_\i$};
        }
        \foreach \n/\i in {s0/$\alpha$, r0/$\beta$}{
        \node[] at ($(\n) + (-4.5mm,0)$) {\i};
        }
        % draw reduced section
        \node[] at ($(r3)!.5!(s3) + (\xOff,0) + (2.8mm,0)$) {$\hat{s}_4$};
    \end{pgfonlayer}
    
\end{tikzpicture}

%% file: fig/forward-estimation/reduced_by_one.tex
\begin{tikzpicture}

    % layers
    \pgfdeclarelayer{nodes}
    \pgfdeclarelayer{edges}
    \pgfdeclarelayer{annotation}
    \pgfsetlayers{edges,nodes,annotation,main}

    \def\xOff{1.5cm} % offset in x between nodes
    \def\yOff{2.7cm} % offset in y between nodes
    \def\nRad{0.8mm} % radius of nodes
    \def\edgeWidth{0.6mm} % width of edges
    \def\pumpRad{3.0mm} % radius of pump symbol
    \def\valveSize{3.0mm} % size of valve symbols 

    \begin{pgfonlayer}{nodes}
        % Return-side nodes
        \node[] (r0) at (0,0) {};
        \node[xshift=\xOff] (r1) at (r0) {};
        \node[xshift=\xOff] (r2) at (r1) {}; 
        \node[xshift=\xOff] (r3) at (r2) {}; 
        
        % Supply-side nodes
        \node[yshift=\yOff] (s0) at (r0) {};
        \node[xshift=\xOff] (s1) at (s0) {};
        \node[xshift=\xOff] (s2) at (s1) {};
        \node[xshift=\xOff] (s3) at (s2) {};

        % draw the nodes
        \foreach \n in {r0, r1, r2, s0, s1, s2}{
        \draw[fill=white] (\n) circle (\nRad);
        }
        % draw valve nodes
        \foreach \r/\s in {r0/s0, r1/s1}{
        \draw[fill=white] ($(\r)!.3!(\s) + (\xOff,0)$) circle (\nRad);
        \draw[fill=white] ($(\r)!.7!(\s) + (\xOff,0)$) circle (\nRad);
        }

    \end{pgfonlayer}

    \begin{pgfonlayer}{edges}
        % connect all the supplies and returns
        \foreach \s/\r in {s1/r1, s2/r2, s3/r3}{
        \draw[black, line width=\edgeWidth] (\s) -- (\r);
        }
        \foreach \si/\sj/\ri/\rj in {s0/s1/r0/r1, s1/s2/r1/r2, s2/s3/r2/r3} {
        \draw[black, line width=\edgeWidth] (\ri) -- (\rj);
        \draw[black, line width=\edgeWidth] (\si) -- (\sj);
        }
        \draw[black, line width=\edgeWidth] (s2) [rounded corners] -- ++(\xOff,0)  -- ($(r2) + (\xOff,0)$) -- (r2);
    \end{pgfonlayer}

    \begin{pgfonlayer}{annotation}
        \foreach \r/\s/\i in {r0/s0/1, r1/s1/2}{
        % draw valve indices
        \node[] at ($(\r)!.5!(\s) + (\xOff,0) + (2.8mm,0)$) {$r_\i$};
        % draw customer pipe indices
        \node[] at ($(\r)!.5!(\s) + (\xOff,0) + (2.8mm,8.0mm)$) {$s_\i$};
        }
        \foreach \s/\i in {s0/5, s1/6}{
        % draw pipe numbers for larger pipes
        \node[] at ($(\s) + (6.6mm, 3.0mm)$) {$s_\i$};
        }
        \foreach \n/\i in {s0/$\alpha$, r0/$\beta$}{
        \node[] at ($(\n) + (-4.5mm,0)$) {\i};
        }
        % draw reduced sections
        \node[] at ($(r2)!.5!(s2) + (\xOff,0) + (2.8mm,0)$) {$\hat{s}_3$};
        
    \end{pgfonlayer}
    
\end{tikzpicture}

%% file: fig/forward-estimation/fully_reduced.tex
\begin{tikzpicture}

    % layers
    \pgfdeclarelayer{nodes}
    \pgfdeclarelayer{edges}
    \pgfdeclarelayer{annotation}
    \pgfsetlayers{edges,nodes,annotation,main}

    \def\xOff{1.5cm} % offset in x between nodes
    \def\yOff{2.7cm} % offset in y between nodes
    \def\nRad{0.8mm} % radius of nodes
    \def\edgeWidth{0.6mm} % width of edges
    \def\pumpRad{3.0mm} % radius of pump symbol
    \def\valveSize{3.0mm} % size of valve symbols 

    \begin{pgfonlayer}{nodes}
        % Return-side nodes
        \node[] (r0) at (0,0) {};
        
        % Supply-side nodes
        \node[yshift=\yOff] (s0) at (r0) {};

        % draw the nodes
        \foreach \n in {r0,s0}{
        \draw[fill=white] (\n) circle (\nRad);
        }

    \end{pgfonlayer}

    \begin{pgfonlayer}{edges}

        \draw[black, line width=\edgeWidth] (s0) -- (r0);

    \end{pgfonlayer}

    \begin{pgfonlayer}{annotation}

        \node[] at ($(r0)!.5!(s0) + (2.8mm,0)$) {$\hat{s}_1$};
    
        \foreach \n/\i in {s0/$\alpha$, r0/$\beta$}{
        \node[] at ($(\n) + (-4.5mm,0)$) {\i};
        }

    \end{pgfonlayer}
    
\end{tikzpicture}